# Multiscale Experiments and Predictive Modelling for Inverse Design and Failure Mitigation in Additively Manufactured Lattices

Mattia Utzeri[a,b], Marco Sasso[b], Vikram S. Deshpande[c], S. Kumar[a,*]

[a]James Watt School of Engineering, University of Glasgow, Glasgow G128QQ, UK

[b]Department of Industrial Engineering and Mathematical Sciences, Polytechnic University of

Marche, Ancona 60121, Italy

[c]Department of Engineering, University of Cambridge, Cambridge CB2 1PZ, UK

**ABSTRACT**

Additive manufacturing (AM) enables the development of high-performance architected cellular materials, emphasizing the growing importance of establishing programmable and predictable energy absorption capabilities. This study evaluates the impact of a precisely tuned fused filament fabrication (FFF) AM process on the energy absorption and failure characteristics of thermoplastic lattice materials through multiscale experiments and predictive modelling. Lattices with four distinct unit cell topologies and three varying relative densities are manufactured, and their in-plane mechanical response under quasi-static compression is measured. Macroscale testing and μ-CT imaging reveal relative density-dependent damage mechanisms and failure modes, prompting the development of a robust predictive modelling framework to capture process-induced performance variation and damage. For lower relative density lattices, an FE model based on the extended Drucker-Prager material model, incorporating Bridgman's correction with crazing failure criteria, accurately captures the crushing response. As lattice density increases, interfacial damage along bead-bead interfaces becomes predominant, necessitating the enrichment of the model with a microscale cohesive zone model to capture interfacial debonding. All proposed models are validated, highlighting inter-bead damage as the primary factor limiting energy absorption performance in FFF-printed lattices. Finally, the predictive modelling introduces an enhancement factor, providing a straightforward approach to assess the influence of the AM process on energy absorption performance, facilitating the inverse design of FFF-printed lattices. This approach enables a critical evaluation of how FFF processes can be improved to achieve the highest attainable performance and mitigate failures in architected cellular materials.

**1. Introduction**

The demand for high-performance materials in industries has driven the need for sustainable solutions. Architected cellular materials are proposed as ideal candidates, as their solid parts are designed to create specific topologies that enhance mechanical or multifunctional performances while ensuring low weight[1–4]. Additive Manufacturing (AM) has played a crucial role in developing innovative cellular materials across length scales, overcoming the limitations of traditional manufacturing technologies and allowing for free-shape design[5–7]. Combining the advantages of AM with high-performance materials has led to cutting-edge applications, including lattices with enhanced structural and/or multifunctional properties[8–11]. Especially, if energy absorption is the main feature to be ensured, the cell topology can be appropriately designed to exhibit a controllable and efficient crushing behaviour[12–16]. Particularly in sectors like aerospace, biomedical, automotive, marine, and defence, there is a growing interest in these cellular materials due to their potential benefits in applications like impact absorbers or crash-worthy constructions[17,18].

In impact applications of this nature, the design of additively manufactured cellular materials relies heavily on the mechanical and thermal characteristics of the base material. Consequently, high-performance polymers, especially Polyetherimide (PEI) like the commercially available Ultem™ 9085, are gaining increased attention. This polymer is typically 3D-printed using Fused Filament Fabrication (FFF) technique. Over time, Ultem™ 9085 has proven its efficacy in various industries due to its amorphous nature, providing excellent adhesive properties. This enables the creation of FFF-printed components with exceptional mechanical, thermal, structural, chemical, electrical, and notably fireproof performance, including flame retardancy, low toxicity, and minimal smoke emission[19]. Extant work primarily focuses on the exploration of FFF-printed PEI, with an emphasis on evaluating the influence of process parameters and deposition strategy on bulk mechanical properties, particularly under tensile and compression loading conditions[20–22]. Recent work by Fores-Garriga et al. has delved into assessing how the unit cell topology influences Young's modulus of PEI 2D and 3D cellular materials[23,24]. However, these studies have predominantly examined the linear elastic properties of PEI cellular materials through numerical, experimental and analytical observations. Notably, the energy absorption characteristics of FFF-printed PEI cellular materials remain unexplored in current research endeavours.

Exploring the mechanics of FFF-printed cellular materials, the predominant focus in this field centres around abundant experimental investigations of mechanical performance[13,25–27]. Studies typically conduct compression tests on architected cellular materials with varying densities and unit cell topologies, evaluating how these factors influence mechanical properties and energy absorption trends. In contrast, the literature features a more limited number of numerical studies capable of predicting the compressive behaviour of FFF-printed cellular materials. These numerical analyses focus on validating FE models applied to lattice structures with various cell topologies. The emphasis is on a specific case of relative density that ensures the robust structural integrity of lattice structures[27,28]. Significantly, there has been a noticeable increase in modelling efforts over the years, corresponding to the maturation of the FFF process. This progress plays a crucial role in elevating the quality of structures printed through FFF, resulting in finely tuned AM lattice materials with minimal variation in mechanical properties. However, challenges arise in establishing a predictive capability for AM-enabled cellular materials and structures using traditional physics-based modelling approaches such as FEM. This is due to the introduction of stochastic/random and systematic/determinate errors inherent in AM processes.

Stochastic errors, stemming from random variations and uncertainties in AM processes, prove challenging to eliminate. Nonetheless, they can be mitigated through process optimization, feedstock and quality control via advanced monitoring and sensing, as well as through modelling and simulation. In contrast, dealing with systematic errors poses a challenge for traditional FEM, hindering the development of a reliable predictive capability. Consequently, recent efforts have shifted towards data-driven mechanics and manufacturing, aiming to develop Artificial Intelligence (AI) and Machine Learning (ML) tools capable of predicting and optimizing the performance of both ordered and disordered cellular materials[29,30]. A recent study by Maurizi et al. utilized an ML-based approach to inversely design lattices with superior buckling performance[31]. While AI and ML tools require substantial data, they may sometimes yield ill-posed or non-physical solutions as they lack a foundation in physics. To address this limitation, recent studies have focused on physics-guided ML tools for predicting and inversely designing AM-enabled lattices[32,33]. For instance, Ha et al. explored inverse design leveraging machine learning to potentially streamline design-manufacturing cycles[34]. In contrast, our study further advances physics-based modelling for the predictive analysis and inverse design of AM-enabled lattices. We posit that the nature of systematic errors lies in inter-bead damage. Consequently, we develop multiscale predictive FE models and scaling laws to evaluate the impact of a well-tuned

FFF AM process on energy absorption and mitigate the failure of PEI lattices, informed by multiscale experiments and characterization.

In glassy thermoplastic polymers like PEI, extensive ductile deformation is primarily governed by shear yielding, while abrupt and brittle fractures are governed by crazing[35,36]. Consequently, the common approach in FE models involves calibrating the Drucker-Prager model, as numerous studies have examined its applicability in different polymers[37–39]. The failure criterion is typically based on the crazing mechanism, following the experimental evidence provided by Sternstein and Ongchin[36]. Contrary to expectations, FFF lattice structures are vulnerable to inter-bead damage, where beads can debond. The existing constitutive models cannot predict such failures, rendering the FE modelling approach generally incapable of providing a comprehensive predictive understanding of the mechanical response of FFF lattice structures. This limitation arises from the coupling of multiple failure modes within lattice structures (intra-layer and inter-layer) with the additive manufacturing process and unit cell sizes[40]. The cell walls of FFF-printed lattice structures, composed of 2, 3, or 4 beads, do not exhibit macroscopic homogeneity, like composite materials[41]. Consequently, conventional debonding criteria in composite mechanics, such as Tsai-Wu, Tsai-Hill, and Puck criteria, cannot be applied. Instead, specific inter-bead failure criteria must be defined along the bead-bead interfaces through cohesive interactions[42–44]. It is essential to note that the mechanics of interfaces and interlayer bonding in FFF-printed components remain actively being explored[40,45]. A couple of recent studies utilise FE constitutive models, based on homogenization techniques, to derive material behaviour from microscale observations[46,47].

Nevertheless, selecting the most suitable additively manufactured lattice structure for industrial applications requires defining programmable mechanical performances through predictive modeling[14,48]. After establishing printing parameters, predicting the mechanical properties of lattice structures becomes crucial, considering the effects of the FFF process. Consequently, this study aims to offer a thorough assessment of the FFF process's impact on the mechanics and energy absorption performances of lattice structures through FE modelling and predictive scaling law approaches. Ultem™ 9085 has been chosen as the base material due to its significance in both industrial and academic contexts. Initially, the in-plane mechanical performance and energy absorption capacity of 3D-printed PEI lattice structures with four distinct unit cell topologies at varying relative densities (20%, 30%, and 40%) were evaluated under monotonic compression. Subsequently, the mechanical responses of FFF-printed lattice structures were analysed through FE modelling, utilizing an extended Drucker-Prager yield criterion incorporating Bridgeman's correction parameter coupled with a crazing failure criterion. This model incorporates the mechanical properties of PEI filament, paired with a failure criterion based on maximum strain calibrated through the Finite Element Model Updating (FEMU) method, ensuring comprehensive consistency with the bulk material response. As the relative density of FFF-printed structures increased, inter-bead damage became predominant. Consequently, the model was enhanced with cohesive interaction along the interfaces of the lattice structures' bead-bead interfaces. In alignment with recent studies on microscale observation, the cohesive interaction was calibrated by leveraging a numerical multiscale approach based on micro-CT results and the Representative Volume Element of bead-bead interfaces. The final stage of predictive modelling involves introducing an enhancement factor, providing a simple scaling law to predict the energy absorption characteristics of FFF-printed structures and estimating the impact of the FFF manufacturing process on the performance and how the FFF AM process can be tuned to mitigate failure in lattice materials.

## 2. Materials and methods

## 2.1. Additive manufacturing of bulk and cellular structures

The Apium P220 Fused Filament Fabrication (FFF) 3D printer was utilized for the fabrication of both lattice structures and fully dense samples using Polyetherimide (PEI) filament feedstocks, commonly referred to as Ultem™ 9085. Supplementary Table 1 provides a summary of all process parameters, which have been meticulously fine-tuned through extensive testing. These parameters collectively optimize the strength, printability and precision of geometric dimensions, ensuring the best outcomes. Consistency in printing parameters has been maintained across all FFF-printed specimens to ensure uniform material properties. To prevent out-of-plane failure, a layer height of 0.1 mm has been selected and the printer's nozzle was equipped with a zone heater located just above the printing surface. Before 3D printing, PEI filament was dried at 60°C for 2 hours.

PEI fully dense (bulk) samples were produced to assess the mechanical properties of FFF-printed PEI under various stress conditions, including tension, compression, bending and triaxial loading. In all bulk samples, the infill density was set at 100% and the longitudinal direction of the bead or layer consistently aligned with the loading direction to ensure that the stress flow direction was in alignment with the orientation of the beads. The PEI lattice structures encompass four distinct unit cell topologies with three varying relative densities ($\bar{\rho}$), representing the solid volume fraction within the cellular structure, at 20%, 30%, and 40%, as illustrated in Fig. 1. Each lattice structure has overall dimensions of 48mm x 48mm x 24mm. To maintain consistent mechanical properties unaffected by boundary effects, the specimens are organized in a 4x4 unit cell array, following methodologies outlined in previous studies[49,50]. The unit cells maintain in-plane dimensions of 12mm x 12mm. To facilitate precise printing of the re-entrant lattice structure with $\bar{\rho} = 20\%$, the lowest cell wall thickness realizable was 0.8 mm. Lattice structures with higher relative densities were designed, leading to an increase in cell wall thickness while retaining the same unit cell size.

The additive manufacturing process achieves this by introducing an appropriate number of beads to attain the desired size for the cell walls. Supplementary Table 2 provides the architectural features for each unit cell topology. It is noteworthy that the S-shaped and I-shaped unit cells are crafted following the anti-chiral strategy, as visually demonstrated in Fig. 1. All geometries were constructed using the nTopology software tool.

## 2.2. Experimental procedures

### 2.2.1. Quasi-static testing

Quasi-static experiments were performed utilizing a Zwick-Roell Z050 universal testing machine. A 5kN load cell was employed to assess the constitutive response of both 3D-printed bulk PEI material and PEI filament, while a 50kN load cell was utilized for compression tests on cellular materials. In all cases, the tests were carried out ensuring quasi-static loading conditions, imposing a strain rate of $10^{-3}$ s$^{-1}$. The lattice structures were compressed between two cylindrical steel plates at a speed of 5 mm/min, and the entire process was recorded using a digital camera. To measure deformations in both tensile and notched tensile tests, a 2D Digital Image Correlation technique was employed[51].

Bulk material properties were evaluated according to ASTM standards. Tensile tests were conducted on FFF-printed dogbone-shaped specimens following ASTM D638. For evaluating flexural characteristics, Three-Point Bending (TPB) tests were performed on FFF-printed rectangular beams, following ASTM D790 guidelines. Compression tests were carried out on FFF-printed cylindrical specimens. Although the standard method (ASTM D695) originally recommended prismatic samples with a length twice their width for compression testing, this led to premature fractures due to compression instability and interlayer delamination. Consequently, a cylindrical specimen with a length equal to its diameter was preferred for these tests[52]. Triaxial properties were assessed using a notched dogbone specimen (ASTM D638) with a triaxiality factor of 0.56. For each configuration, including bulk samples, PEI filament and cellular materials, three specimens were tested to ensure the repeatability of experimental results.

*2.2.2. Differential scanning calorimetry (DSC)*

The glass transition temperatures ($T_g$) were analyzed in three different scenarios: one for the PEI filament (before printing), one for the bead of the FFF-printed dogbone (post-printing), and one for the bead of the re-entrant lattice structure with relative density ($\bar{\rho}$) of 20% (post-printing). The glass transition temperatures of the beads and filament were determined using Differential Scanning Calorimetry (DSC) analysis using the DSC Seiko Exstar 6000 instrument. Samples were heated at a rate of 10 ºC/min from room temperature (25 ºC) to 250 ºC with a constant nitrogen flow of 50 mL/min. The $T_g$ values were calculated based on the midpoint between extrapolated heat flow.

*2.2.3. Micro-computed tomography (µCT)*

The extent of porosity and its distribution in the as-printed condition of lattice structures bulk samples were assessed through micro-computed tomography (µCT) analysis. The 2D lattice structures were scanned after the compression tests to highlight the damage mechanisms associated with in-plane crushing. The cross-sectional images of the cellular materials and bulk samples were captured using Metrotom Tomography (ZEISS) instrument, achieving a voxel size of about 24 µm. The acquired scans were reconstructed and analysed using Matlab software.

**2.3. Multiscale Finite element modelling**

The study involved a series of Finite Element (FE) analyses, including the validation of the mechanical behaviour of both bulk (fully dense) and cellular materials and the estimation of failure parameters for both the bulk material and the bead-bead interface. Consequently, the numerical characteristics of the FE models are summarized below.

*Dogbone FE model*: A plane-stress state was assumed to replicate the tensile response of the dogbone specimen, given its thinness and the observed planar fracture in physical tensile tests. A 4-node bilinear plane stress quadrilateral element was chosen with reduced integration and hourglass control (CPS4R). An average mesh size of 0.05 mm was utilized for the analysis. *Three-Point Bending (TPB) FE Model:* To simulate the TPB beam's behaviour, a plane-strain constraint was applied to the longitudinal section of the TPB beam due to its width being approximately 12 times its thickness, thereby meeting the conditions for a plane-strain state. A 4-node bilinear quadrilateral element for plane strain analysis, featuring reduced integration and hourglass control, known as the CPE4R element, was chosen. The model was configured with a mesh size of 0.05 mm. *Cylindrical FE model*: To simulate the compressive response of the cylindrical sample, an axisymmetric condition was imposed. A 4-node bilinear axisymmetric quadrilateral element with reduced integration and hourglass control, referred to as the CAX4R element, was selected for the analysis. The model was created with an average mesh size of 0.05 mm. *Notched Dogbone FE Model*: The notched dogbone-shaped specimen's FE model was constructed to capture triaxiality properties. This FE model exhibits the same numerical characteristics as the Dogbone FE model, apart from an average mesh size of 0.01 mm, particularly in proximity to the notch tip. *RVE-B FE Model*: For the 3D FE model of the RVE-B, which includes bead-bead interfaces, an 8-node linear brick element with reduced integration and hourglass control (C3D8R) was employed for meshing. Periodic Boundary Conditions (PBCs) were applied to the external faces[53]. The overall mesh size was set at 0.01 mm, while the edges at the bead-bead interface comprised an average mesh size of 0.001 mm. *Lattice FE models*: The planar cross-section of each type of lattice structure was meshed using CPE4R elements due to the large width of the lattice structures ensuring a plane strain condition. The mesh size was set at 0.1 mm. The compressive plates were represented as rigid wires within the FE model. The wire length exceeded structure sizes to ensure uniform compression even if the structure lost its integrity. The numerical analyses on *Lattice FE models* were performed using the Abaqus/Explicit solver whereas the other FE models utilized the Abaqus/Implicit solver. The contact interactions were defined with normal and tangential contact behaviours, incorporating friction properties based on a penalty factor of 0.1 and contact separation following compression loading (hard contact). These contact interactions were applied to all surfaces, including newly created surfaces resulting from fracture.

## 3. Constitutive and damage models

### 3.1 Drucker-Prager model with Bridgman's correction (D-P model with BC)

Constitutive models and failure criteria for glassy thermoplastic materials such as PEI or PMMA usually take into account the competition between shear-yielding and crazing[35]. To represent shear-yielding, we employ the extended Drucker-Prager yield criterion which shows substantial material ductility without fracture. To model crazing, we introduce a craze-initiation criterion based on the local maximum principal strain which depends on the local mean stress. Continuum constitutive relation for craze widening is not considered, thus the craze-breakdown and failure coincide with the craze-initiation criterion. In addition, the isotropic elastic regime before the initiation of plastic deformation was considered. After the yielding point, the work hardening was associated with the true stress-strain curve obtained from a uniaxial tensile test conducted on the PEI filament. The following sections delve into the description of the models and the calibration of the models' parameters. Note that the initially considered *D-P model* is recalled in Sections 4.2 and 4.6 to highlight its inadequacy to capture the flexural response of the parent material and compression response of lattice structures. The *D-P model* incorporates the work hardening associated with the true stress-strain curve obtained from a uniaxial compression test on an FFF-printed cylinder and the crazing failure criterion was calibrated with the ultimate failure strain measured for FFF-printed dogbone samples, as presented in Section 4.1. Therefore, the *D-P model* might be considered as the constitutive model calibrated following the traditional approach.

*3.1.1 Shear-yielding*

Recent studies show that, especially for glassy thermoplastic polymers, the extended *D-P* model provides accurate prediction for determining pressure-dependent yield criterion[38]. The extended *D-P* model postulates the second invariant of the deviatoric stress tensor $S_{ij}$, denoted as $J_2$, follows a linear combination with the first invariant, $I_1$, of the Cauchy stress tensor $\sigma_{ij}$:

$$J_2 = A + BI_1 \quad (1)$$

where *A* and *B* are material constants and

$$I_1 = \sigma_{kk} = 3\sigma_m; \quad (k = 1, 2, 3) \quad (2)$$

$$J_2 = \frac{1}{2} S_{ij} S_{ij} \quad (3)$$

$$S_{ij} = \sigma_{ij} - \frac{\sigma_{kk}}{3} \delta_{ij} \quad (4)$$

where $\delta_{ij}$ is the Kronecker delta and $\sigma_m$ is the mean stress. The extended *D-P* constitutive model is implemented in Abaqus. To calibrate the yield criterion, we used both uniaxial tension and compression yield strengths of FFF-printed bulk samples, presented in Section 4.1. In addition, a non-associated flow rule with the dilation angle set close to zero was chosen, ensuring incompressible inelastic deformation, thereby preserving the convexity of the yield surface and eliminating any dependency on the third deviatoric stress invariant[27]. The calibrated model parameters are listed in Table 1.

*3.1.2 Bridgman's Correction*

For FE modelling of large strain problems like the present one, it is imperative to comprehend the post-necking behaviour of PEI filament to precisely capture the strain hardening and real flow curve. Therefore, the true stress-strain response of the PEI filament from the onset of necking was determined following the G'Sell approach[54], as the filament displayed necking as evidenced by the load-displacement curve depicted in Figs. 3a and S1. The true stress-strain relationship was calculated up to the onset of necking (corresponding to peak load) using traditional continuum mechanics theory, given the homogeneous deformation. Beyond the peak load, the necking region experiences both radial and circumferential stresses, leading to the loss of the uniaxial stress condition. The G'Sell approach provides a framework to obtain the material's true stress-strain curve from the deformation map of the necked axisymmetric region, based on Bridgman's correction parameter. A comprehensive explanation of this procedure is provided in the Supplementary Note 1.

*3.1.3 Crazing*

Bowden and Oxborough developed an empirical crazing-initiation criterion based on critical strain[35], given by

$$\varepsilon_1 = \frac{1}{E}\left[C(t,T) + \frac{D(t,T)}{\sigma_1 + \sigma_2 + \sigma_3}\right] \qquad (5)$$

where $\varepsilon_1$ is the maximum principal strain and $E$ is the polymer Young's modulus. $\sigma_i$ denotes principal stresses (where $i = 1, 2, 3$) and both $C(t,T)$ and $D(t,T)$ are time- and temperature-dependent material constants. Assuming craze-initiation is independent of time and temperature, Equation 5 was rearranged, introducing the first invariant, $I_1$, of the Cauchy stress tensor $\sigma_{ij}$, as follows:

$$\varepsilon_1 = X + \frac{Y}{I_1} \qquad (6)$$

where *X* and *Y* are material constants. Crazing widening was considered negligible because craze-breakdown was assumed to occur at very low localized strain. To support this assumption, Gearing calibrated the complete crazing model showing a craze-breakdown strain of approximately 0.005[35]. Consequently, polymer fracture can be straightforwardly modelled through Equation 6. This failure criterion was implemented in Abaqus through a user-defined subroutine utilizing the element-removal technique. The failure criterion was calibrated through inverse identification according to the Finite Element Model Updating (FEMU) method[55]. In the FEMU procedure, the dogbone (tensile) and TPB (flexural) FE models were used to simulate the real experiments. A cost function $\Phi(X,Y)$ is defined, which encompasses the mismatch between numerical and experimental results:

$$\Phi(X,Y) = \phi_D + \phi_{TPB} = \left|\frac{\varepsilon_f^{Exp} - \varepsilon_f^{Fem}}{\varepsilon_f^{Exp}}\right| + \left|\frac{\delta_f^{Exp} - \delta_f^{Fem}}{\delta_f^{Exp}}\right| \qquad (7)$$

where $\phi_D$ and $\phi_{TPB}$ are the Normalized Root Mean Square Deviations (NRMSD) associated with tensile and flexural loading conditions, respectively. Here, $\varepsilon_f^{Fem}$ and $\varepsilon_f^{Exp}$ are the numerical (*Dogbone FE model*) and experimental ultimate strains measured and estimated over the dogbone gauge length respectively. $\delta_f^{Fem}$ and $\delta_f^{Exp}$ are the numerical (*TPB FE model*) and experimental ultimate displacements, respectively, of the beam at the mid-span. The material constants *X* and *Y* are iteratively updated until the cost function is minimized below a given threshold. The minimization procedure was carried out using Matlab software employing the optimization solver "Fmincon". Figure S2 depicts the flowchart of the optimization procedure. The converged *X* and *Y* values by inverse identification are summarized in Table 2 and the crazing failure criterion is graphically shown in Fig. 4a.

### 3.2. Drucker-Prager model with Bridgman's correction and debonding (D-P model with BC & Debonding)

The *D-P model with BC & Debonding* enriches the *D-P model with BC* with supplementary traction-separation cohesive law to mimic localized damage occurring along the interfaces between adjacent beads (referred to as interlayer debonding, as depicted in Fig. 7). Notably, the *D-P model with BC* falls short in predicting interlayer damage, as it assumes that the cell walls are uniform and possess homogeneous properties. To capture interlayer damage, the *Lattice FE models* were updated. The planar shapes of lattice structures were subdivided based on the deposition strategy outlined in G-code instructions, forming cell walls with a specified number of beads/layers (see Supplementary Fig. 3). The interfacial damage/facture between beads/layers within the cell walls of the complete lattice structure is captured through a Cohesive Zone Model (CZM) described below.

*3.2.1. Cohesive zone modelling*

Cohesive Zone Modelling allows for the representation of interlaminar delamination phenomena through the establishment of a traction-separation model. Traction-separation model assumes initially linear elastic behaviour defined through penalty factors $K_n$ and $K_s$ followed by the damage initiation and damage evolution regime. In this context, the quadratic nominal stress damage initiation criterion is employed, signifying that the bead-bead interface experiences damage when:

$$\left(\frac{\langle t_n \rangle}{t_n^{max}}\right)^2 + \left(\frac{\langle t_s \rangle}{t_s^{max}}\right)^2 = 1 \tag{8}$$

where the $t_n$ is the normal traction (normal to the interface) and $t_s$ is the in-plane shear stress. $t_n^{max}$ and the $t_s^{max}$ are normal and shear strengths. The damage evolution model describes the degradation of cohesive interaction using a damage variable $D$ which evolves from 0 to 1 upon further loading after the initiation of damage:

$$t_n = \begin{cases} (1-D)\,\bar{t}_n, & \bar{t}_n > 0 \\ \bar{t}_n, & \bar{t}_n < 0 \end{cases} \tag{9}$$

$$t_s = (1 - D)\,\bar{t}_s$$

where $\bar{t}_n$ and $\bar{t}_s$ are the contact stress components predicted by the elastic traction-separation behavior for the current separations without damage. The damage evolution was defined based on energy release rate and fracture toughness, so the mixed-mode criterion is implemented:

$$\frac{G_I}{G_{TC}} + \frac{G_{II}}{G_{SC}} = 1 \qquad (10)$$

where $G_I$ is the energy release rate corresponding to normal traction and $G_{II}$ is the energy corresponding to in-plane shear stress. $G_{TC}$ and $G_{SC}$ are the maximum fracture energy in normal traction and shear modes respectively and $G_C = G_{TC} + G_{SC}$. The damage variable $D$ is obtained from energy information as follows:

$$D = \frac{\delta_m^f(\delta_m^{max} - \delta_m^0)}{\delta_m^{max}(\delta_m^f - \delta_m^0)} \qquad (11)$$

where the effective separation at complete failure is $\delta_m^f = \frac{2G_c}{t_{eff}^0}$ in which $t_{eff}^0$ is the effective traction at damage initiation. $\delta_m^0$ is the effective separation at damage initiation and $\delta_m^{max}$ is the maximum value of the effective separation attained during the loading history. The traction-separation law was implemented in Abaqus as a surface-based cohesive interaction.

The cohesive model was calibrated by leveraging the Representative Volume Element (RVE) theory: the RVE for Bead interfaces (RVE-B) mimics the mechanical behaviour of the bead-bead interface. Therefore, a 3D geometric model of the bead-bead interface was built from a dogbone µCT scan and then imported into Abaqus for numerical simulation (*RVE-B FE model*). The RVE-B employs the *D-P model with* BC assuming homogeneous properties inside the bead. In fact, from a microstructural point of view, amorphous polymers, like PEI, have great polymer chain diffusion, meaning polymer chains can fuse across layers exhibiting near-isotropic properties[40]. The intralayer damage was estimated based on the maximum normal and shear stresses that the RVE-B can withstand. Fig. 2 shows the flow chart and summarises the strategy employed to predict the bead-bead interface properties.

The CZM was calibrated using the traction-separation response predicted by numerical testing on RVE-B. For example, the traction response of RVE-B depicted in Fig. 2b was obtained by stretching it along the traction, $t_n$ up to the fracture. The normal mode separation $\delta_n$ was measured as the difference between the closest fully solid cross-sections, including the porosity shape before and during the loading (see Supplementary Fig. 4). The normal traction was computed as the load divided by the RVE-B external area normal to $t_n$. The penalty factor $K_n$ defines the slope at the early stage of traction-separation response as shown in Fig. 2d, $K_n$ = 26000 N/mm³. Following Turon et al., the total fracture energy was estimated as the area under the traction-separation curve[56], $G_{TC}$= 0.22 mJ/mm². The maximum normal traction was defined as the load at which the failure occurred, $t_n^{max}$ = 58 MPa. The in-plane shear properties were defined as well: $t_s^{max}$ =34 MPa, $G_{SC}$ = 0.86 mJ/mm², and $K_s$ = 2150 N/mm³. To qualitatively assess the maximum traction and shear predictions generated by numerical testing on the RVE-B, a comparison with existing literature is conducted. The attained results reveal that the $t_n^{max}$ is approximately 70% of the bulk material strength where beads are aligned with loading direction, while the $t_s^{max}$ accounts for around 40% of the same benchmark. Notably, a few studies indicate that the transverse strength of bulk materials denoted here as $t_n^{max}$, is approximately 70% - 80% of the tensile strength of bulk materials with beads aligned to the loading direction. Furthermore, the shear over tensile strength experiences a reduction of 50% according to findings from various studies[40,43,57]. Consequently, the implemented numerical procedure provides reliable estimations.

**4. Results and discussion**

**4.1 Experiments: mechanical response of PEI filament and FFF-printed specimens**

The quasi-static mechanical behaviour of the FFF-printed PEI bulk sample exhibits distinct characteristics: it displays a brittle response when subjected to tensile loading conditions (as depicted in Supplementary Fig. 5a), whereas it shows a ductile behaviour under compression loading, as shown in Supplementary Fig. 5b. The ductility of FFF-printed PEI is further confirmed by its flexural response, as depicted in Supplementary Fig. 5c. The compressive yield stress, $\sigma_{CS}$ is estimated to be 70 MPa, while the tensile yield stress, $\sigma_{TS}$ is estimated to be 62 MPa. Notably, Young's modulus in tension and compression, $E_S$ is consistent and measures 2150 MPa. Through DIC analysis, the Poisson's ratio is determined to be 0.3596 and the fracture strain measured as the average true strain in the gauge length, $\varepsilon_f$ is found to be about 0.05.

The PEI filament exhibits necking, as illustrated in the load-displacement response shown in Fig. 3a during tensile loading. This behaviour is a typical trait observed in glassy thermoplastic polymers, resulting from the reorganization of polymer chains as they align along the loading direction[54]. Ultimately, the polymer undergoes brittle failure once all polymer chains are fully oriented. As detailed in the preceding section, the true stress-strain ($\sigma_t$–$\varepsilon_t$) relationship can be determined using G'Sell's procedure (Supplementary Note 1 for more details). Beyond the peak stress, the $\sigma_t$–$\varepsilon_t$ relationship for the PEI filament follows a power law strain-hardening pattern as illustrated in Fig. 3b, with a hardening exponent of approximately 3. PEI filament exhibits a tensile strength of about 175 MPa with a failure strain of 0.65, in stark contrast to the brittle response evidenced by the tensile test on the FFF-printed dogbone specimen.

### 4.2 FE validation of PEI bulk properties

Before conducting FE simulations on lattice structures, the mechanical response of FFF-printed PEI specimens under tensile, compression, flexural and triaxial loading was validated, utilizing the *D-P model with BC*. It is noteworthy to highlight the FE validation of the tensile and flexural responses of 3D-printed specimens. Notably, the failure criterion in the *D-P model with BC* predicts a higher fracture strain under uniaxial tensile loading than the experimentally measured value - 0.23 vs. 0.05 (red dot vs. dotted vertical line in Fig. 4a and Fig. 4c, respectively). Despite this apparent contradiction with the brittleness observed in FFF-printed dogbone specimens, the *D-P model with BC* accurately predicts dogbone response, confirming its effectiveness and consistency, as shown in Fig. 4c. To understand the cause, attention is directed to the FE simulation. After surpassing the yielding point (Figure 4d), the FE model predicted necking, akin to the behaviour of PEI filament in a tensile test, leading to a subsequent shear band. Concurrently, localized strain surpassed the fracture strain (0.23), preventing the attainment of a stable cross-section. Despite this, the average true strain within the gauge length remained around 0.05, concealing the abrupt localization of strain. The shear band coincided with the location of the fracture in the FFF-printed dogbone specimen as shown in Fig. 4d. Once failure occurred, the fracture propagated normally to the loading direction, consistent with theoretical and experimental expectations. Similar fracture behaviour is commonly observed in other studies on FFF-printed dogbone specimens[42,58]. Note that if the failure strain were higher, a stable neck region would extend along the specimen's axis, mirroring the behaviour observed with the PEI filament.

Figure 4b compares FE predictions with experimental results from the TBP test. The experimental data highlights a significant mismatch in mid-span displacement, denoted as δ, in stark contrast with *D-P model*. This discrepancy arises because *D-P model* was calibrated using the ultimate strain exhibited by the dogbone sample, specifically the average true strain in the gauge length (0.05). *D-P model with BC* rectifies this issue and accurately replicates the experimental observations. It can be concluded that FEMU provides $\varepsilon_f$ values aligning with experimental data and effectively captures the ductility of FFF-printed PEI, previously concealed by shear banding in dogbone specimens. Furthermore, the *D-P model with BC* also validates notched tensile and compression tests, aligning well with the experimental results, as displayed in Figs S6 and S7.

### 4.3. Differential scanning calorimetry analysis

The Differential Scanning Calorimetry (DSC) analysis, as illustrated in Supplementary Fig. 8, indicates that the glass transition temperature, $T_g$ of the PEI filament is approximately 181.2°C, a value consistent with Ultem 9085™. There were no significant differences in $T_g$ between the beads forming the lattice structure and the dogbone, both measuring at approximately 177°C. Notably, the $T_g$ of FFF-printed PEI decreased by approximately 3% due to the FFF processing. Changes in $T_g$ can be indicative of variations in mechanical properties. Turner demonstrated an empirical relationship between fracture strength, $\sigma_f$ and molar mass, $M$ for polymers, inspired by Flory's equation, given by the formula[59]:

$$\sigma_f = \sigma_\infty - \frac{A_s}{M} \tag{12}$$

Here $\sigma_\infty$ is the polymer strength with infinite molar mass and $A_s$ is a material constant. The molar mass is related to the polymer $T_g$ through the Fox-Flory equation:

$$T_g = T_{g\infty} - \frac{B_s}{M} \tag{13}$$

where $T_{g\infty}$ is the polymer $T_g$ with infinite molar mass and $B_s$ is a material constant. Novikov et al. demonstrated the Fox-Flory relationship for several glassy thermoplastic polymers[60]. Given the change in $T_g$ measured through DSC and Equations 12 and 13, it could be hypothesized that the polymer strength decreases due to FFF processing, consequently leading to a decrease in fracture strain. This observation was also confirmed by FEMU inverse identification, which validates the tensile response of FFF-printed PEI by estimating a lower fracture strain compared to the PEI filament - 0.23 vs. 0.65. Therefore, it could be inferred that the FFF additive manufacturing process significantly affects the material's fracture strain, altering it from before to after the 3D printing process. On the contrary, the true stress-strain response might be assumed to resemble the filament one, as the *D-P model with BC* validates various stress states (tension, compression, flexural, and triaxial) by employing the true stress-strain curve of the PEI filament.

### 4.4. Experiments: quasi-static compression of PEI lattices

The in-plane compression behaviour of all 3D-printed PEI 2D lattices with three different relative densities was measured under quasi-static loading conditions. The deformation and failure patterns of lattices with $\bar{\rho} = 20\%$ captured at different stages of loading along with macroscopic stress-strain response are shown in Fig. 5: A (Initial collapse stress - $\sigma_p$), B (Onset strain of densification - $\varepsilon_d$), and C (Densification). The stress-strain curves exhibit three well-known regimes: an initial elastic phase, a plateau regime in which lattices exhibit stable or unstable tress fluctuations and the densification phase. The in-plane compressive response of the structures turns out to be dependent on the unit cell topology[61]. The re-entrant lattice shows a stretching-dominated behaviour exhibiting an extended linear elastic regime followed by a sudden drop in stress. The re-entrant structure collapse is governed by the cell wall buckling phenomenon and extended plastic deformation which creates a stable compressive response leading to layer-by-layer collapse up to near complete densifaction[62]. The hexagonal, I-shaped and S-shaped lattice structures show a bending-dominated behaviour, exhibiting a nonlinear elastic regime followed by fractures or extended plastic deformation. Specifically, the hexagonal lattice structure exhibits an elastic regime followed by layer-by-layer collapse, forming crush bands and ensuring high plateau stress. The auxetic I-shaped and S-shaped lattice structures show a strong nonlinear elastic response because of the formation of additional load transfer paths between adjacent cell walls, as shown in Fig. 5 (see, stage A). Note that the S-shaped structure shows a transition from a bending-dominated to a stretching-dominated mechanical response once the ligaments are completely packed. The fracture of vertical and horizontal struts which connect the "S" shaped features leads to a brittle crushing behaviour in the plateau regime typical of stretch-dominated structures[61]. On the contrary, the local rotation of the ligaments in the I-shaped structure turns out to be less critical than the S-shaped structure, giving it a stable compressive response in the plateau regime and leading to smooth densification.

The mechanical response of PEI lattice structures changes as their relative density increases. The stress-strain curves start to exhibit a progressive brittle crushing regime dominated by ligament fractures and preferential damage along the bead-bead interfaces, i.e., interlayer damage (inter-bead debonding) as shown in Fig. 7. Such a mechanical response is evident in Fig. 6 in which the in-plane quasi-static compression behaviour of PEI lattice structures with $\bar{\rho} = 40\%$ is shown. The interlayer/inter-bead damage is also predominant at the intermediate relative density $\bar{\rho} = 30\%$ as shown in Supplementary Fig. 9. The interlayer failures are quite noticeable in the hexagonal lattice structure at stage A where the initial collapse occurs due to debonding of bead-bead interfaces. As the platen compacts the structure, the debonded beads separate and warp independently reducing the integrity of cells.

The energy absorption performance of the FFF-printed lattice structures is experimentally evaluated by comparing the onset of densification strain $\varepsilon_d$, the initial collapse stresses $\sigma_p$, and the specific energy absorption (SEA) as shown in Supplementary Fig. 10 and Table 3. The $\sigma_p$ is the peak stress reached by the structure before the collapse. The energy absorbed by the lattice structure per unit volume is defined as follows:

$$W = \int_0^{\varepsilon_d} \sigma \, d\varepsilon \quad (14)$$

and the SEA is expressed as follows:

$$SEA = \frac{W}{\rho} = \frac{1}{\rho} \int_0^{\varepsilon_d} \sigma \, d\varepsilon \quad (15)$$

where $\rho$ is the average density of the lattice structure. Table 3 shows that the highest energy absorption capacity of FFF-printed structures with $\bar{\rho} = 20\%$ was achieved by the hexagonal lattice structures, including the highest $\sigma_p$ and $\varepsilon_d$. Although the $\sigma_p$ slightly increases as the relative density increases the SEA does not follow the same trend. That means the increase in relative density does not allow the hexagonal structure to reach higher energy absorption properties as happens for all FFF-printed structures. It is well-known that the SEA of lattice structure increases as the relative density increases in both stretching-dominated and bending-dominated lattice structures[2,63]. This correlation holds under the condition that the ligaments are defect-free, meaning they are without interlayer damage[58,59]. Consequently, if there are defects present in the cell walls in their as-processed condition, the monotonic increase in SEA with the increase in relative density might not be established.

### 4.5. Micro-computed tomography analysis

Micro-computed tomography (μCT) imaging was conducted on the lattice structures before (as-printed) and after undergoing compression tests. The re-entrant structure (as-printed) with a relative density, $\bar{\rho}$ =20%, has no intra-bead or inter-bead micro-pores and defects, as illustrated in Fig. 7a. The μCT image of the as-printed lattice structure is unable to differentiate between beads within the cell walls, where each cell wall comprises two beads, each with a thickness of 0.4 mm. This feature guarantees robust structural integrity and further validates the choice of 0.8 mm as the minimum thickness for printable cell walls. In the case of the as-printed re-entrant lattice structure with $\bar{\rho} = 40\%$, porosities were primarily observed along the bead-bead interfaces and near the cell edges, as depicted in Fig. 7c. Notably, porosities were more pronounced at the interfaces between printed beads/layers than within the beads, mainly due to the elliptical cross-section of the beads, as emphasized by several studies[40,43,64]. The porosity sizes near the cell edges remained consistent along the Z-direction, as it is dependent on the extrusion path. The overall structural porosity in the lattice structure, as revealed by µCT analysis, was approximately 8.5%, resulting in a real relative density slightly below the intended value – 0.37 compared to the designed 0.4. The µCT analysis confirmed the accuracy of the structural dimensions, with the measured thickness of each bead at 0.41 mm compared to the designed 0.4 mm (see Supplementary Fig. 3). This ensures that the mechanical properties obtained through experimental testing consistently pertain to a cellular material with the correct architectural parameters. Consequently, the potential for unforeseen variations in mechanical performance is predominantly associated with porosity, rather than geometric imperfections or their coexistence.

Figure 7d highlights two distinct damage mechanisms: red circles denote cracks propagating transversely to the cell walls (crazing failure), while blue ellipses highlight interlayer/inter-bead damage, specifically debonded beads within the cell walls. During compression loading, failure at the interfaces between beads leads to bead disbonding, causing warping, separation, and debonding similar to delamination in laminated materials - described as interlayer debonding. Lower-density structures do not exhibit interlayer debonding; instead, cracks propagate transversely to the beads or cell walls, as shown in Fig. 7b. With increasing relative density, re-entrant structures display both damage mechanisms (see, Fig. 7d) due to structural weakening at the bead-bead interfaces, as depicted in Fig. 7c. Volume reconstruction shown in Supplementary Fig. 11 reveals no signs of out-of-plane fractures, confirming strong adhesion between stacked layers for $\bar{\rho} = 20\%$ and predominant in-plane interlayer failure for $\bar{\rho} = 40\%$. This ensures the impact of the FFF process, evaluated in this study, on the mechanical performance of lattice structures is only connected to in-plane inter-bead damage. The distribution of voids within the bulk (fully dense) samples was also evaluated and presented in Supplementary Fig. 12, with the bulk sample porosity measuring 3.6%.

## 4.6. FE validation of PEI lattice structures

The numerical predictions of the in-plane compression behaviour of re-entrant lattice structure with $\bar{\rho} = 20\%$ are compared with the experimental results in Fig. 8 (see Supplementary Movie 1), showing the characteristic engineering stress-strain curves, deformation maps, and failures at various stages. Both the *D-P model* and the *D-P model with BC* effectively capture the compressive response, accurately predicting the buckling phenomena of the central unit cell layer up to the initial collapse. However, the *D-P model* exhibits brittle fracture once the structure buckles (Figure 8, stage A), followed by a subsequent plateau regime governed by brittle crushing mechanisms (Figure 8, stages B and C). In contrast, the *D-P model with BC* predicts a buckled shape of a re-entrant cell that can be completely folded, mirroring the experimental test (Figure 8, stage A). The *D-P model with BC* accurately captures the layer-by-layer buckling phenomena, compacting cells without fractures in the plateau regime. The sequential buckling initiates from the central cell layer and progresses towards the cells near the compressive plates, aligning with the experimental findings observed in stages B and C. The ductility exhibited in stretched regions plays a pivotal role in the post-buckling response of the re-entrant structure, confirming the simultaneous presence of plastic yielding and elastic buckling mechanisms in the plateau regime. Consequently, the *D-P model with BC* perfectly validates the experimental compressive response of the re-entrant structure with $\bar{\rho} = 20\%$, confirming again the accuracy of the *D-P model with BC* versus the *D-P model*.

As the relative density of FFF-printed structures increases, predictions of the *D-P model with BC* become increasingly inaccurate because it cannot capture the inter-bead/interlayer debonding occurring at higher relative density as shown in Fig. 9. The interlayer debonding changes the compressive response of the FFF-printed structures, weakening the overall mechanical performance as described in the previous section. However, the *D-P model with BC & Debonding* accurately captures the in-plane compressive response of higher relative density lattice structures as it employs a CZM to capture the interfacial damage between beads. The numerical predictions of in-plane compression behaviour of re-entrant lattice structure with $\bar{\rho} = 40\%$ are compared with the experimental results in Fig. 9 (see Supplementary Movie 2), showing the characteristic engineering stress-strain curves, deformation maps and failures at various stages. The *D-P model with BC* predicts the buckling of the central cell layer and subsequent layer-by-layer brittle fractures (see, Fig. 9 at stages B and C). On the contrary, the experimental stress-strain curve exhibits a smooth transition from the elastic to plateau regime coupled with hardening up to the first fracture, showing an initial collapse stress 2.5 times lower than that predicted by the *D-P model with BC*. Re-entrant cell buckling cannot occur at this stress level because the load is lower than the critical load required for triggering buckling instability and the initial collapse stress cannot be connected to cell buckling in turn. Structures with plateau regimes based on plastic collapse mechanisms usually exhibit similar mechanical behaviour, which means inhomogeneous deformations inside the re-entrant cell walls occur[27,49]. The FFF-printed PEI material does not reach the yielding point even if thick cell walls experience a higher multi-axial stress state[65]. Thus, the inhomogeneous deformations are limited to the bead-bead interfaces, weakening the structural integrity of the re-entrant lattice structure. The macroscopic collapse response of re-entrant structure changes from layer-by-layer to diagonal (green ellipses in Fig. 9 at B stage) and the *D-P model with BC & Debonding* was able to predict the macroscopic collapse. In addition, the model validates the extensive bending of the beads close to the cell edges confirming the beads' ductility.

Note the change in slope from the elastic to plateau regimes is due to the progressive failure of cohesive interaction. Once the interfaces along the structure diagonals are broken, the collapse of re-entrant cells occurs. A comparison of the predicted and experimental compressive response of hexagonal lattice structure with $\bar{\rho} = 30\%$ is provided in Supplementary Note 6 and in Supplementary Fig. 13.

*4.6.1. Parametric study*

A systematic FE parametric study was conducted on all four different lattice structures with three different relative densities using both the *D-P model with BC* and the *D-P model with BC & Debonding*. A comprehensive comparison of numerical results with those of experiments is presented in the form of engineering stress-strain curves in Fig. 10. In addition to Fig. 10, complete numerical validation of the in-plane compression behaviour of PEI lattice structures is shown in Supplementary Fig. 14 ($\bar{\rho} = 20\%$) and Supplementary Fig. 15 ($\bar{\rho} = 40\%$). All three figures demonstrate that while the *D-P model with BC* can predict accurately the stress-strain response for lower relative density lattice structures, it can't accurately capture the crushing behaviour of high relative density structures. However, the *D-P model with BC & Debonding* captures the macroscopic crushing response of all lattices tested across three different relative densities accurately, showing consistent deformed shapes, stress-strain behaviour and energy absorption characteristics. The numerical results further confirm that inter-bead damage plays a fundamental role in the mechanical response of FFF-printed lattice materials.

**4.7 Enhancement factor**

The previous section explored numerical and experimental results emphasizing the influence of interlayer damage in as-printed lattices on the in-plane compressive response of FFF-printed lattice structures. The overall mechanical performance is diminished by the interlayer damage found in lattices with high relative density. Mechanical properties, including energy absorption characteristics, generally rely on relative density. For instance, the energy absorption capacity (EAC) of lattice structures can be expressed as:

$$W = C\left(\frac{\rho}{\rho_s}\right)^\alpha \tag{16}$$

where $C$ and $\alpha$ are topology-dependent constants[63]. The value of $\alpha$ for lattice structures typically ranges from 1.5 to 2, while 3D lattice structures can exhibit values up to 3[15,62]. Figure 11a reveals that the FFF-printed PEI lattice structures have an $\alpha$ parameter close to 1, indicating that the EAC of FFF-printed lattice structures deviates significantly from theoretical values. Theoretical estimate assumes defect-free cell walls: if the solid part of cellular material, like in FFF-printed structures, has inter-bead defects within the cell walls, Equation 16 provides a lower EAC compared to that of the same cellular structure with defect-free cell walls.

The estimation of performance decline can be achieved by introducing an enhancement factor, denoted as $\mu$, which establishes a connection between the EAC of a perfect lattice structure and a flawed one, as per the approach outlined by Gibson and Ashby[66]. Since the interlayer damage, as well as geometric imperfections, are closely related to the FFF process, the enhancement factor $\mu$ is defined as follows:

$$\frac{W}{W^*} = \mu \tag{17}$$

Here, $W$ represents the EAC of a lattice structure fabricated via FFF and $W^*$ denotes the EAC of the same ideal structure i.e., without interlayer damage. Analogously, in the context of Gibson and Ashby theory, $W^*$ corresponds to the EAC of a perfect lattice structure. The value of $W^*$ is obtained using the *D-P model with BC*, assuming homogeneous constitutive properties within the cell walls (absence of interlayer damage) of the lattice structure. Indeed, Fig. 11a illustrates that the *D-P model with BC* predicts a relationship between $W$ and $\bar{\rho}$ in a log-log graph according to theoretical estimate ($\alpha \simeq 2$).

The enhancement factor $\mu$ serves as a straightforward metric for assessing the impact of the AM process on EAC. However, $\mu$ must be defined as a key parameter of the FFF printing strategy, effectively linking it to the cell wall thickness to bead thickness ratio ($t/d_n$), where $t$ is the cell wall thickness, and $d_n$ is the nozzle diameter. Figure 11b demonstrates that $\mu$ decreases with an increase in $t/d_n$ ratio, indicating a higher number of beads within the cell walls. This leads to a decline in lattice structure performance due to an elevated degree of bead-bead interfacial defects and damage. A best-fitting equation for $\mu$ is provided as:

$$\mu = -\frac{t}{4\,d_n} + 1.5 \qquad (18)$$

The slope of the enhancement factor line is closely linked to process parameters, with a lower slope for the perfect lattice structure (horizontal line with a $\mu = 1$) indicating higher interlayer strength. Achieving the best performance involves minimizing the area between the dotted slant line and the solid horizontal line (representing $\mu = 1$) in Fig. 11b. In other words, reducing the shaded area denoted by $\Pi$ to nearly zero results in fully solid cell walls with homogeneous material properties and intact bead-bead interfaces, indicating the highest print quality for FFF-printed structures. It is noteworthy that Habib et al. explored the mechanical properties of FFF-printed lattice structures across a range of relative densities, achieving theoretical values[67]. However, they consistently ensured the presence of two beads across the thickness of cell walls by adjusting the nozzle size, maintaining $t/d_n$ ratio close to 2. This practice ensures the highest mechanical performance attainable by FFF-printed lattice structures and mitigates failure.

**5. Conclusion**

The study elucidates the effects of a meticulously tuned fused filament fabrication (FFF) additive manufacturing (AM) process on the energy absorption capabilities and failure behaviour of 2D polyetherimide (PEI) lattices. Employing a comprehensive approach that integrates multiscale experiments and predictive modelling, we specifically examine the in-plane crushing behaviour and failure mechanisms of FFF-printed PEI lattice structures, incorporating four distinct unit cell topologies and three relative densities. At lower relative densities, we observe anticipated bending- or stretching-dominated deformation behaviour in the plateau regime. However, as relative density increases, the compressive responses undergo a shift in deformation and damage mechanisms. Low relative density structures exhibit homogeneous deformation within cell walls, while high relative density lattice structures manifest preferential damage along bead-bead interfaces (interlayer damage), as evidenced by µ-CT scans. Consequently, high relative density lattices demonstrate insufficient improvements in energy absorption performance due to their susceptibility to interlayer debonding. Our multiscale finite element models further corroborate the detrimental impact of interlayer damage on failure behaviour and energy absorption performance.

The *D-P model*, initially calibrated through monoaxial tests, presents inconsistent flexural responses of the parent material and compression responses of lattice structures. As a remedy, we introduce an alternative model, the *D-P model with BC*, which accounts for the influence of the FFF process. This model utilizes the mechanical properties of PEI filament corrected by Bridgman's correction parameter and calibrates the crazing failure criteria through inverse analysis, aligning with experimental observations on FFF-printed dogbone and three-point bending specimens. Successfully capturing tensile, flexural, triaxial and compression behaviours of FFF-printed parent material, as well as the crushing response of low relative density lattice structures, the enhanced model falls short in accurately representing the crushing response of high relative density lattice structures where inter-bead damage prevails. To address this limitation, we augment the *D-P model with BC* by introducing cohesive interactions along bead-bead interfaces (*D-P model with BC & Debonding*). This enhancement, informed by numerical simulations of a microscale RVE comprising bead-bead interfaces, effectively validates the compressive response of FFF-printed high relative density lattices and underscores the detrimental role of inter-bead damage.

In the final phase of our predictive modelling, we introduce a scaling law based on the enhancement factor ($\mu$) and the cell wall thickness to bead thickness ratio. This straightforward method offers insights into the impact of the FFF AM process on the energy absorption and failure characteristics of FFF-printed lattices. The enhancement factor provides clarity on the quality of FFF-printed structures and the efficacy of chosen process parameters. Notably, maintaining a cell wall thickness-to-bead thickness ratio close to 2 proves crucial for optimal mechanical performance, steering clear of interlayer damage and ensuring a high-quality FFF-printed structure. While intralayer damage poses challenges to the structural integrity of FFF-printed structures, the introduction of the enhancement factor serves as a practical tool for predicting and evaluating the impact of FFF-3D printing on the mechanical performance and failure behaviour of FFF-printed structures. Moreover, the enhancement factor extends the utility of the well-established predictive scaling laws proposed by Gibson and Ashby, providing a flexible parameter for the inverse design of additively manufactured cellular materials and enhancing the adaptability and precision of the design process.

**Supplementary Information**

Supplementary information includes tables comprising architectural parameters of lattice structures & process parameters, G'Sell approach to obtain true stress-strain response of PEI filament based on Bridgman's correction parameter, inverse identification procedure by FEMU, micro-CT analysis of cellular & bulk materials, DSC analysis, FE validation of the compression & triaxial responses of 3D printed parent/bulk material, experimentally evaluated compressive behaviour of PEI lattice structures with $\bar{\rho} = 30\%$, FE validation of in-plane compressive behaviour of PEI lattice structures with $\bar{\rho} = 20\%$ & $\bar{\rho} = 40\%$, synchronized video (Supplementary Movie 1) of stress-strain response & deformation map of experimental vs numerical re-entrant lattice structure with $\bar{\rho} = 20\%$ and synchronized video (Supplementary Movie 2) of stress-strain response and deformation map of experimental vs numerical re-entrant lattice structure with $\bar{\rho} = 40\%$.

**Acknowledgements**

Financed by the European Union-NextGenerationEU (National Sustainable Mobility Center CN00000023, Italian Ministry of University and Research Decree No.1033-17/06/2022, Spoke 11- Innovative Materials & Lightweighting), and National Recovery and Resilience Plan (NRRP), Mission 04 Component 2 Investment 1.5-NextGenerationEU, call for tender n. 3277 dated 30 December 2021. The authors express their gratitude to Alejandro Tray and Johannes Schneider for their valuable assistance in the 3D printing process.

**Author Contributions**

S.K. conceived the idea. S.K. and V.S.D. supervised the project. M.U. performed 3D printing, experiments, and numerical analyses. M.S. supported the implementation of user-defined subroutine and the inverse identification approach. All authors contributed to the preparation of the manuscript and S. K. derived the key insights.

**Competing interests**

Authors declare no financial and non-financial competing interests in the subject matter or materials discussed in this article. The opinions expressed are those of the authors only and should not be considered representative of the European Union or the European Commission's official

**Table of Contents**

**Graphical abstract**

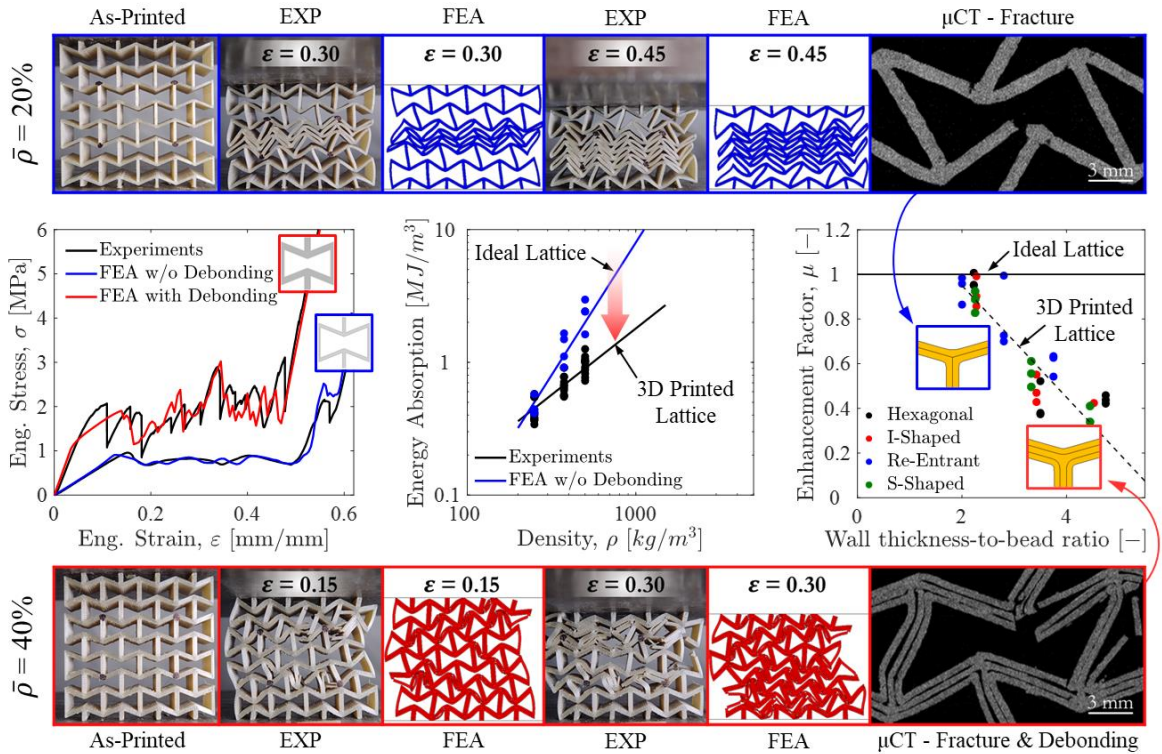

**Figures**

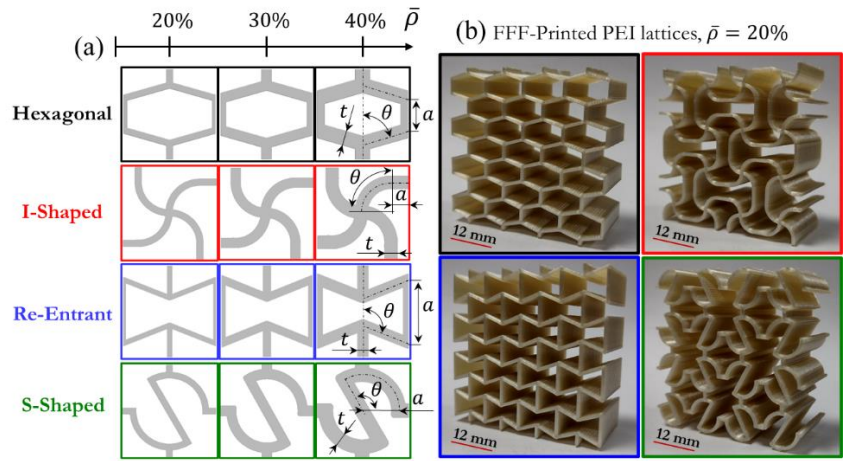

Fig. 1: Two-dimensional lattice structures: (a) Geometric models for different relative densities illustrating architectural parameters of various unit cell topologies and (b) additively manufactured Polyetherimide (PEI) lattice structures for a relative density $\bar{\rho} = 20\%$, featuring Hexagonal (top-left), I-shaped (top-right), Re-Entrant (bottom-left), and S-shaped (bottom-right) configurations. Details of architectural parameters are summarized in Supplementary Table 2.

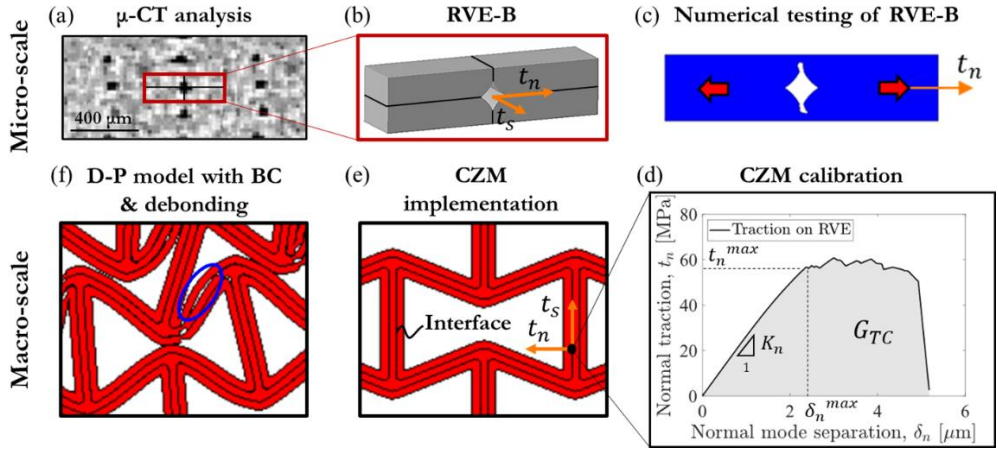

Fig. 2: Flowchart showing the steps involved in modelling the macroscale compression behaviour of FFF-printed lattice structures considering inter-bead failure: (a) μ-CT image showing interfaces between layers and beads topology. (b) Building RVE-B from microscale observation. (c) FE modelling RVE-B. (d) Traction-separation response obtained from numerical tensile test on RVE-B. (e) CZM implementation along bead-bead interfaces to capture inter-bead debonding. (f) Debonding response predicted by *D-P model with BC & debonding*

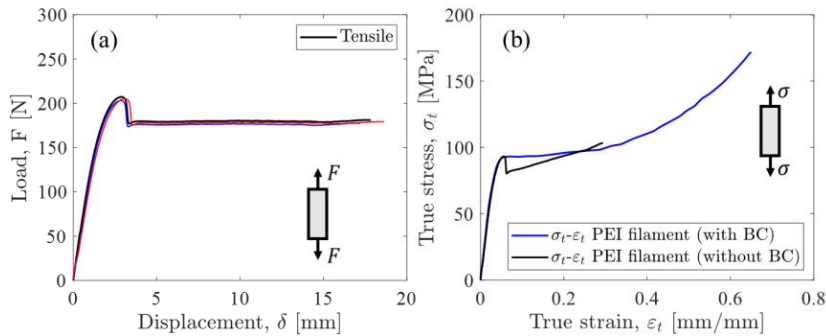

Fig. 3: Quasi-static tensile response of PEI filament: (a) Load-displacement behaviour and (b) true stress-strain curve with and without considering Bridgman's correction (BC).

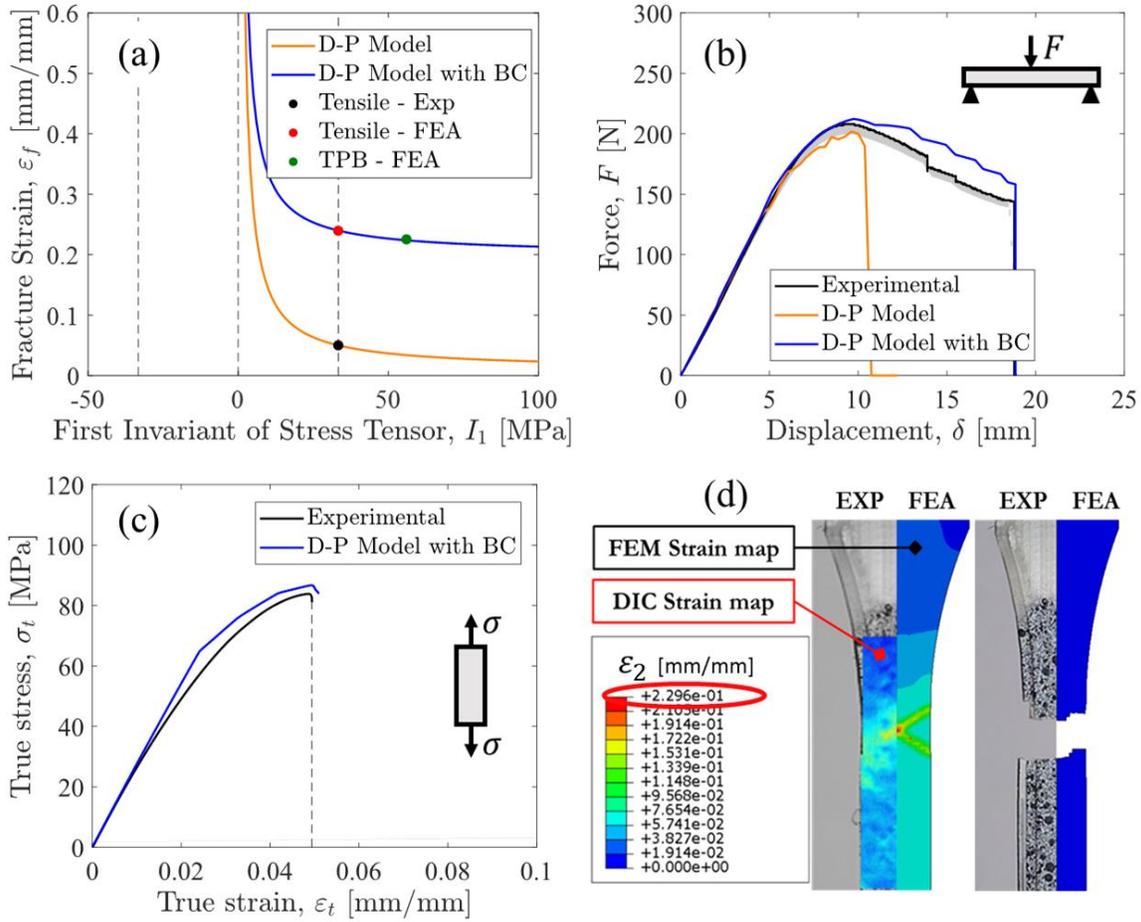

Fig. 4: FE validation of parent material properties under tension and three-point bending (TPB). (a) Crazing failure criterion calibrated in D-P model with BC and D-P model. Dots highlight the fracture strain obtained from tensile and TPB tests. (b) Experimental vs numerical results of FFF-printed PEI specimen under flexural loading. (c) Experimental vs numerical results of FFF-printed PEI specimen under tensile loading. (d) DIC vs FEA strain maps and experimental vs FEA failures of FFF-printed PEI dogbone.

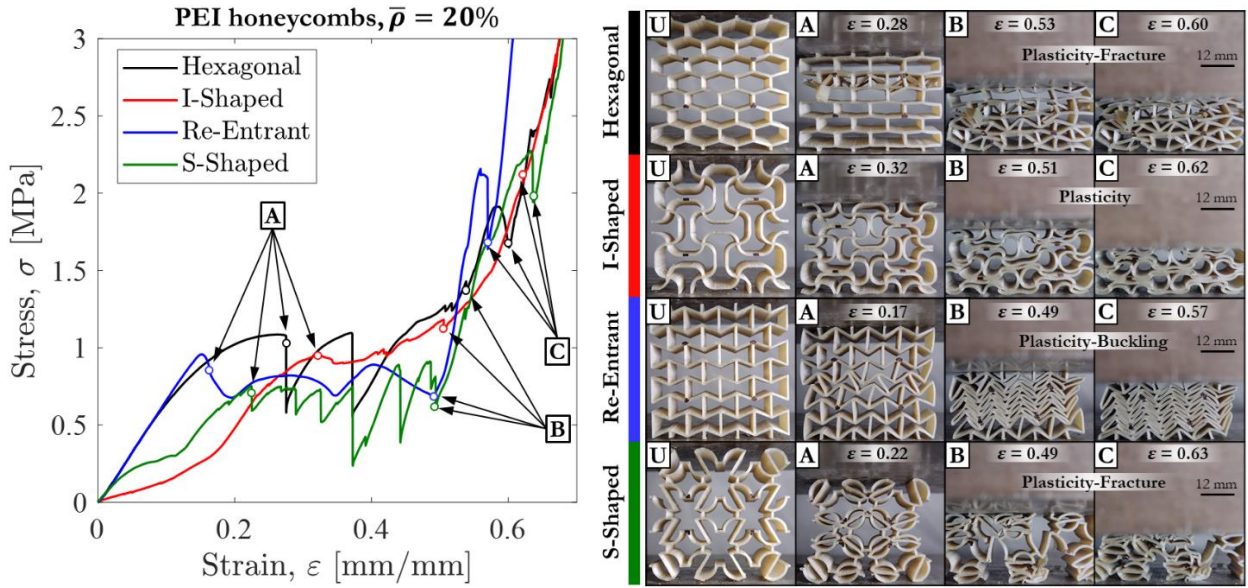

Fig. 5: In-plane quasi-static compression behaviour of PEI 2D lattices with $\bar{\rho}$= 20%. Characteristic engineering stress-strain response and deformation maps at various stages including failure modes are shown: U (Undeformed), A (Initial collapse stress), B (Onset of densification), and C (Densification).

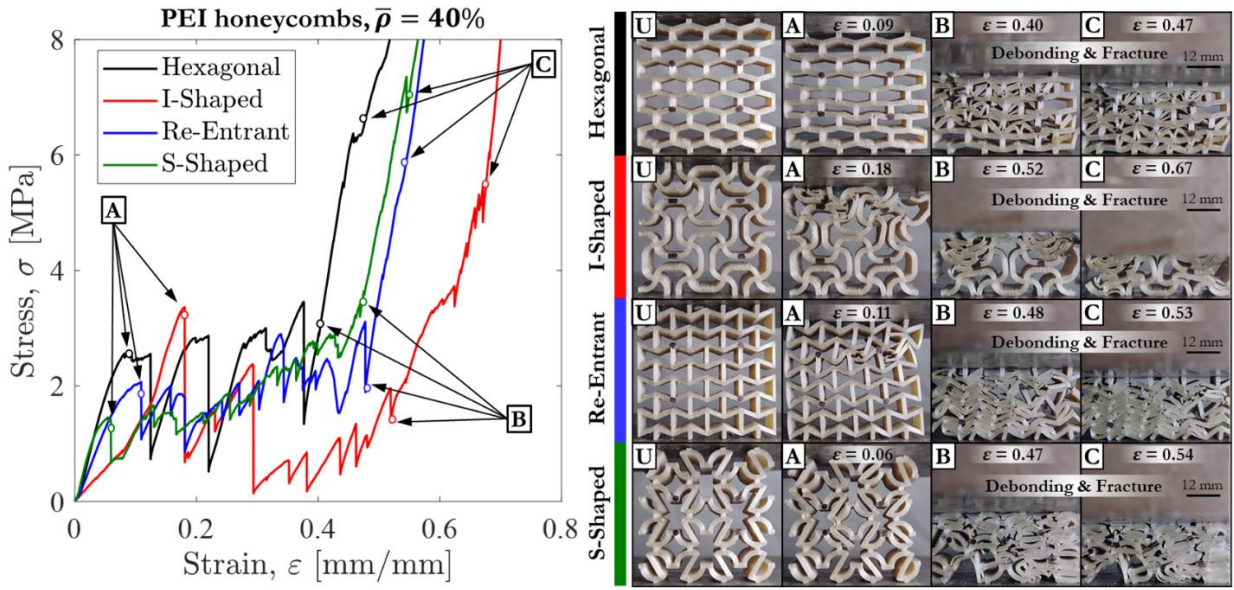

Fig. 6: In-plane quasi-static compression behaviour of PEI 2D lattices with $\bar{\rho}$ = 40%. Characteristic engineering stress-strain response and deformation maps at various stages including failure modes are shown: U (Undeformed), A (Initial collapse stress), and B (Onset of densification), and C (Densification).

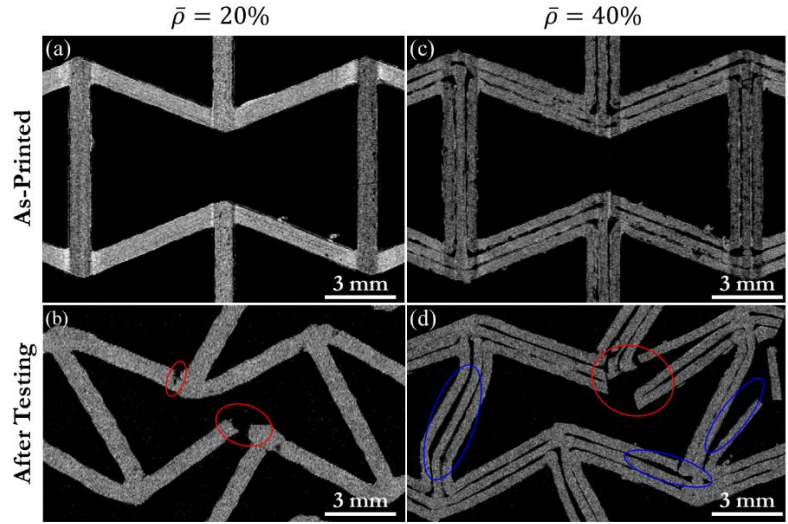

Fig. 7: Micro-Computed Tomography (μ-CT) analysis of re-entrant lattice structures: (a) $\bar{\rho}$ = 20% - before compression tests. (b) $\bar{\rho}$ = 20% - after compression tests, (c) $\bar{\rho}$ = 40% - before compression tests and (d) $\bar{\rho}$ = 40% - after compression tests.

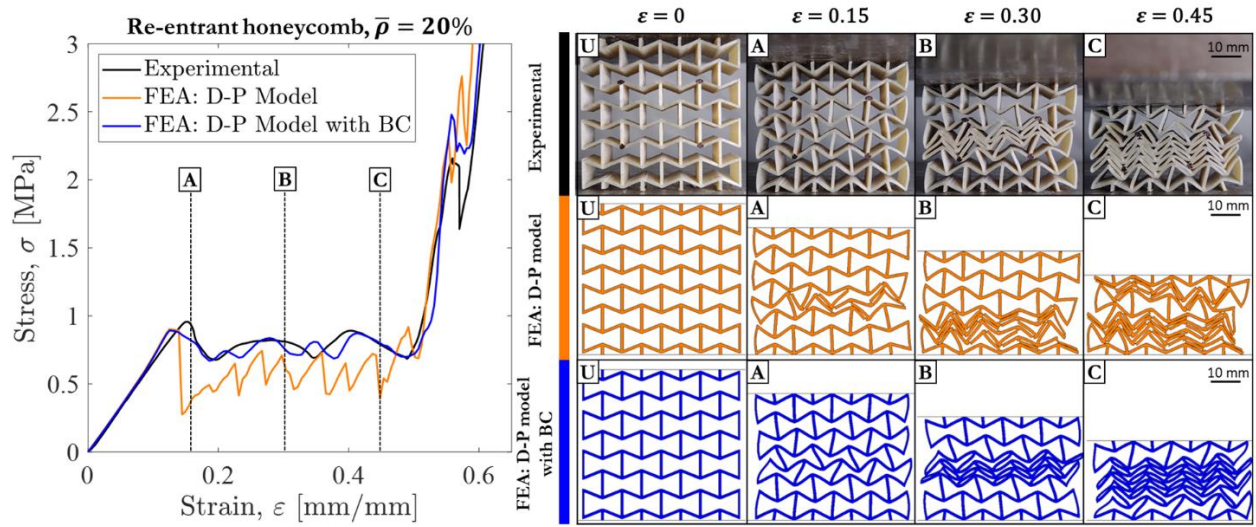

Fig. 8: Experimental vs FE predictions of in-plane compression behaviour of re-entrant lattices with $\bar{\rho}$ = 20%. Characteristic engineering stress-strain response and deformation maps at various stages including failure modes are shown: U (Undeformed), A (ε = 15%), B (ε = 30%), and C (ε = 45%). (see Supplementary Movie 1)

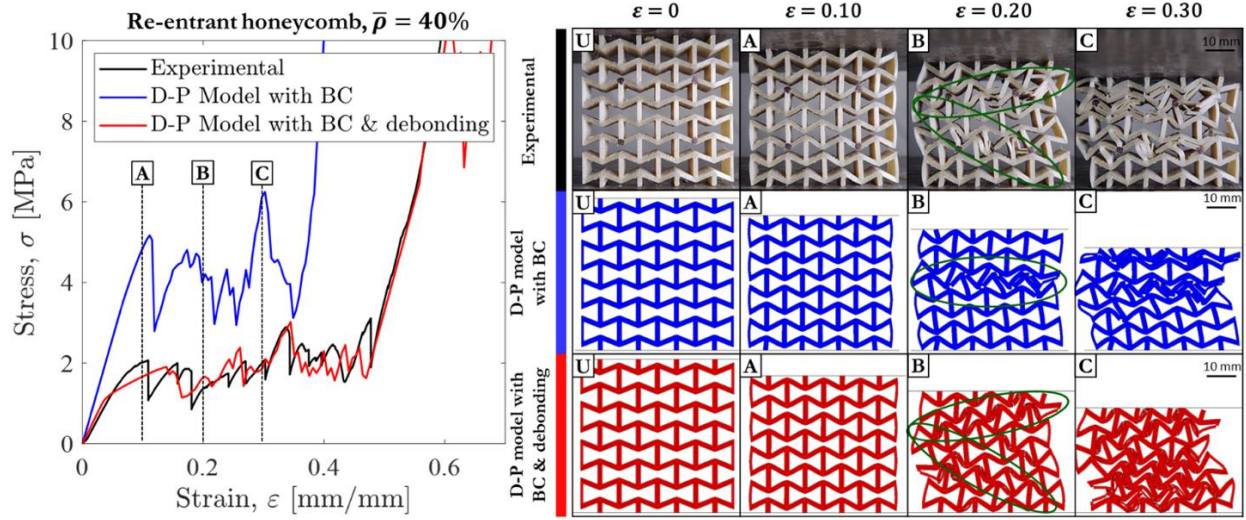

Fig. 9: Experimental vs FE predictions of in-plane compression behaviour of re-entrant lattices with $\bar{\rho}$ = 40%. Characteristic engineering stress-strain response and deformation maps at various stages including failure modes are shown: U (Undeformed), A ($\varepsilon$ = 10%), B ($\varepsilon$ = 20%), and C ($\varepsilon$ = 30%). (see Supplementary Movie 2)

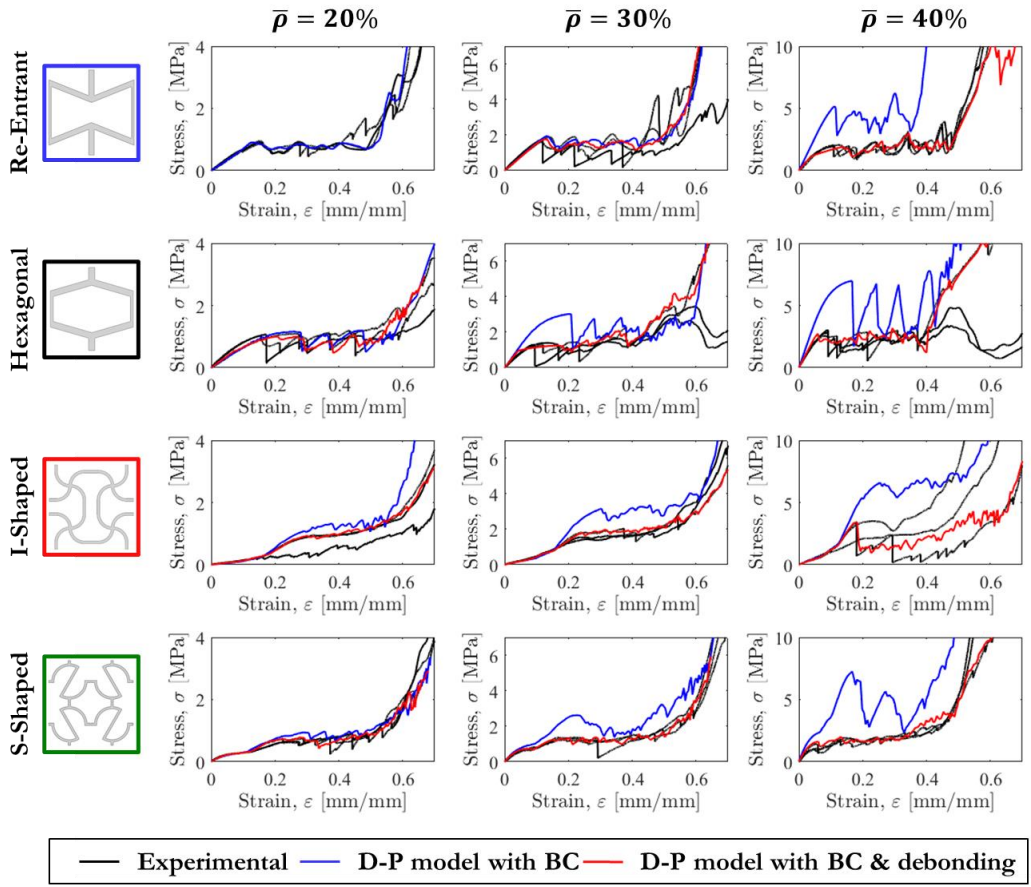

Fig. 10: Comparison of in-plane compression behaviour in PEI honeycombs: engineering stress–strain curves from experimental and numerical analyses under quasi-static compression.

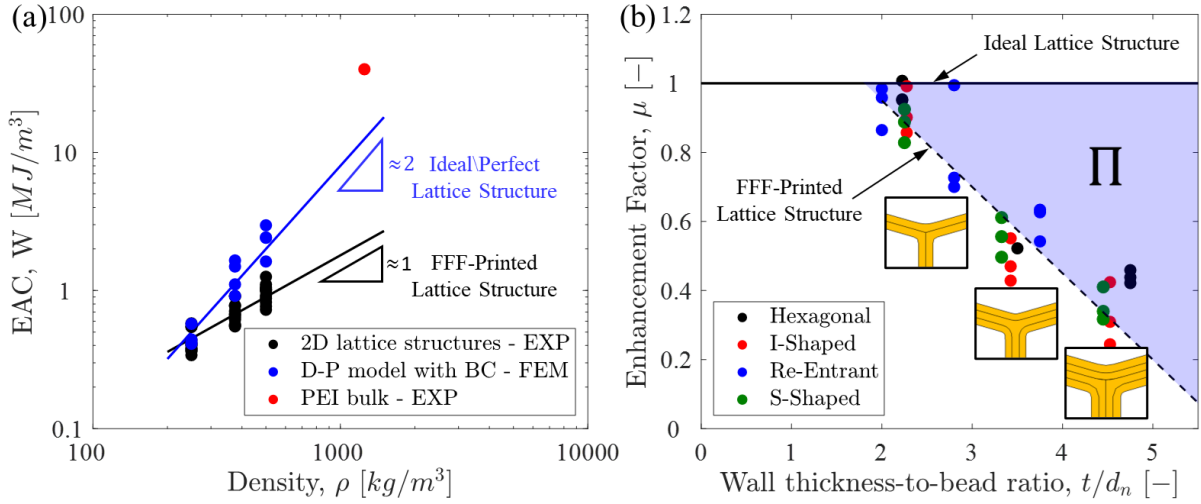

Fig. 11: Impact of FFF process on energy absorption performances of PEI lattice structures. (a) Influence of inter-bead damage on the energy absorption capacity of FFF-printed PEI lattice structures. (b) Predictive energy absorption scaling law in FFF-printed lattice structures based on enhancement factor and wall thickness-to-bead ratio.

**Tables**

Table 1: Calibrated parameters for extended Drucker-Prager plasticity model

| A [MPa$^2$] | B [MPa] | Dilation angle [deg] | Flow potential eccentricity |
|---|---|---|---|
| 1466.66 | -2.66 | 1 | 0.1 (Default) |

Table 2: Calibrated parameters for crazing failure criterion

| FEA model | X [mm/mm] | Y [MPa mm/mm] |
|---|---|---|
| *D-P model* | 0 | 1.332 |
| *D-P model with BC* | 0.201 | 1.301 |

Table 3: Summary of in-plane compression performances of FFF-printed PEI lattices.

| Unit cell topology | $\bar{\rho}$ [%] | E [MPa] | $\nu$ [-] | $\sigma_p$ [MPa] | $\varepsilon_d$ [mm/mm] | SEA [J/g] |
|---|---|---|---|---|---|---|
| Hexagonal | 20 | 7.22±0.6 | 0.28±0.01 | 1.09±0.02 | 0.53±0.03 | 2.19±0.12 |
| Hexagonal | 30 | 22.56±2.5 | 0.27±0.02 | 1.28±0.06 | 0.44±0.02 | 1.73±0.15 |
| Hexagonal | 40 | 41.43±4.6 | 0.26±0.02 | 2.63±0.21 | 0.40±0.01 | 2.11±0.09 |
| I-Shape | 20 | 1.21±0.3 | -0.83±0.02 | 0.96±0.21 | 0.51±0.08 | 1.52±0.13 |
| I-Shape | 30 | 4.14±1.2 | -0.77±0.03 | 1.84±0.23 | 0.54±0.02 | 2.43±0.29 |
| I-Shape | 40 | 10.32±3.4 | -0.81±0.02 | 3.34±0.50 | 0.52±0.06 | 1.45±0.58 |
| Re-Entrant | 20 | 7.26±0.6 | -0.38±0.02 | 0.95±0.03 | 0.49±0.03 | 1.67±0.08 |
| Re-Entrant | 30 | 15.93±1.8 | -0.31±0.01 | 1.91±0.14 | 0.55±0.04 | 2.56±0.54 |
| Re-Entrant | 40 | 28.08±4.3 | -0.36±0.02 | 2.06±0.11 | 0.48±0.02 | 2.06±0.10 |
| S-Shape | 20 | 5.08±0.5 | -2.17±0.02 | 0.75±0.07 | 0.49±0.01 | 1.27±0.09 |
| S-Shape | 30 | 12.71±1.9 | -1.73±0.04 | 1.33±0.05 | 0.54±0.05 | 1.81±0.16 |
| S-Shape | 40 | 37.85±3.6 | -1.48±0.02 | 1.47±0.17 | 0.47±0.04 | 1.99±0.26 |

Supplementary Information for

**Multiscale Experiments and Predictive Modelling for Inverse Design and Failure Mitigation in Additively Manufactured Lattices**

Mattia Utzeri[a,b], Marco Sasso[b], Vikram S. Deshpande[c], S. Kumar[a,*]

[a]James Watt School of Engineering, University of Glasgow, Glasgow G128QQ, UK

[b]Department of Industrial Engineering and Mathematical Sciences, Polytechnic University of Marche, Ancona 60121, Italy

[c]Department of Engineering, University of Cambridge, Cambridge CB2 1PZ, UK

**Supplementary Note 1. G'Sell approach and Bridgman's Correction**

The PEI filament exhibits consistent necking behaviour during the tensile test. The localized deformation occurring in the necking region complicates the conventional method of deriving a stress-strain curve from a tensile test on PEI filament. G'Sell suggests an alternative approach for extracting an accurate stress-strain curve ($\sigma_t - \varepsilon_t$) within the axisymmetric necked region using load-displacement data[1]. Supplementary Fig. 1 illustrates the outlined strategy for obtaining the corrected PEI material response. Starting from the load drop point in the load-displacement curve, as depicted in Supplementary Fig. 1a, the PEI filament exhibits stable neck propagation along the gauge length, a phenomenon commonly observed in polymers. For instance, Supplementary Fig. 1b depicts the necked region during deformation stage B. Photographic analysis of the necked region was conducted to ascertain the filament radius ($R$) using the optical image of the necked area. The image underwent post-processing, transitioning from grayscale to a binarized format, to achieve the appropriate contrast for identifying the filament profile and subsequently determining the filament radius along the filament $y$-axis. Once the filament radius is defined, the associated curvature $k$ is defined according to a well-known equation: $\frac{1}{k} = \frac{\partial^2 R}{\partial y^2}$. The uniaxial true strain at the necked region is commonly measured as $\varepsilon_t = 2ln\left(\frac{R_0}{R}\right)$, where $R_0$ is the radius of the filament in undeformed configuration. On the contrary, the uniaxial stress needed further correction. Notably, Bridgman identified the curvature radius as the key parameter to predict the effective monoaxial properties of material in the necked region. Bridgman thus introduced an effective stress ($\sigma_{eff}$) which is numerically equal to the uniaxial stress which would produce the same effective strain as the one generated by the complex stress field, for axisymmetric specimens:

$$\sigma_{eff} = \frac{\sigma_y}{\zeta_B}$$

(1)

where $\sigma_y = \frac{P}{\pi R^2}$ is the measured stress and $\zeta_B$ is the correction parameter proposed by Bridgman, commonly known as Bridgman's Correction (BC) parameter[2] given as:

$$\zeta_B = \left(1 + \frac{2k}{R}\right) ln\left(1 + \frac{R}{2k}\right)$$

(2)

$\zeta_B = 1$ when the profile is uniform and $\zeta_B \neq 1$ in the necking region. However, Bridgman defined this correction after several tests on metals in which the necked region is not stable. On the contrary, in polymer, the necked region stable and goes from one uniform profile to another uniform profile as shown in Supplementary Fig. 1b. For this reason, G'Sell investigated the effectiveness of BC also in the polymer necked region and concluded that the BC is a good first approximation for the triaxiality effect. Therefore, the BC could be used also for polymer necked region.

Before the necking (blue section in Supplementary Fig. 1c), $\sigma_{eff} = \sigma_y$ because the profile is uniform, i.e. $\zeta_B = 1$ as displayed in Supplementary Fig. 1c. Running from the blue section to the green section in Supplementary Fig. 1d, the profile is convex due to the radial stress being negative (compressive state on stretched filament). Therefore, the $\sigma_{eff}$ should be higher than $\sigma_y$ to obtain the same value of uniaxial tensile strain, as confirmed by Supplementary Fig. 1d. The shape of the PEI filament profile leads to a rapid increase of $\sigma_{eff}$ close to the blue section getting a value comparable with the peak of stress measured before the drop and becomes quite stable as the strain increases. Running from the green section to the orange section in Supplementary Fig. 1d, the profile is concave so the $\sigma_{eff}$ is lower than $\sigma_y$ (tensile state on stretched filament). Away from the necked region, $\sigma_y = \sigma_{eff}$ because the profile is uniform, i.e. $\zeta_B = 1$. In the orange section, the $\varepsilon_t$ is 0.48. To conclude, the true stress-strain relationship of PEI filament was calculated up to the onset of necking (corresponding to peak load) using traditional continuum mechanics theory, given the homogeneous deformation. Beyond the peak load, the procedure described above was repeated for several pictures up to the PEI filament failure, building true stress-strain curve of PEI filament from the obtained $\sigma_{eff} - \varepsilon_t$ curves, as shown in Supplementary Fig. 1e.

**Supplementary Note 2. Inverse identification procedure**

The inverse identification procedure aims to define the unknown $X$ and $Y$ parameters of crazing failure criteria. The inverse identification was based on FEMU, and the entire procedure was automatized using Abaqus and Matlab scripts. A Matlab script was used to write a USDFLD subroutine in Fortran that manages the element removal/deletion according to the $X$ and $Y$ updating variables. Then, Matlab script enables the numerical simulation of *FE dogbone* and the *FE TPB models,* integrating the USDFLD subroutine. Once the simulations were completed, the results were exported, and the cost function was calculated. The minimization algorithm 'Fmincon' was used to update the design variables obtaining the lowest values of the cost function $\Phi(X, Y)$. When the cost function reaches a stable value below 5% the minimization procedure ends.

**Supplementary Note 3. Micro-Computed Tomography analysis**

*Supplementary Note 3.1. Porosity distribution*

Micro-CT scans were analysed to investigate the porosity distribution within PEI parent materials printed in different shapes. As depicted in Supplementary Fig. 12, beads exhibit alignment with the deposition strategy and G-Code paths, forming an orderly pattern. The voids, situated between beads at the corners of the beads are due to the elliptical cross-section of beads and can be quantified as the count of voids within the cross-section. In Supplementary Fig. 12, voids exhibit air density, and the grayscale values correspondingly represent air content. The process involves converting the grayscale image to a binarized image, followed by measuring the number of voxels associated with the solid part. The effective solid part is determined by the ratio of the measured solid zone to the expected one, i.e., the cross-section measured with a calliper. The porosity value is then derived as the complementary value. In this instance, dogbone samples reveal a porosity of 3.06%. The same methodology was applied for calculating porosity in lattice structures.

*Supplementary Note 3.2. Bead-bead interfaces*

The interfaces between beads become apparent in FFF-printed structures due to the presence of voids along the edges of stacked beads. Consequently, the absence of material in these regions causes the cell walls of lattice structures to appear divided. Supplementary Fig. 3 provides a clearer illustration of this concept through a comparison between the volume reconstruction of a re-entrant lattice structure, as per the CAD model and CT scans. The dimensions of the beads are closely tied to FFF printing parameters, including nozzle diameter, layer height, extrusion multiplier and printing overlap. Nevertheless, once the FFF printing parameters are established, the beads maintain consistent shapes and sizes across all printed samples. In this study, the beads have a width of 0.41 mm and their shape is depicted in Supplementary Fig. 3.

*Supplementary Note 3.3. Out-of-plane failure of re-entrant lattice structures*

Supplementary Fig. 11, utilizing CT scans, illustrates the robust out-of-plane structural integrity of FFF-printed lattice structures. No indications of cracks, debonding, or disbonding between stacked layers are observed. This observation affirms that any inter-bead damage is confined solely to the thickness direction of the cell walls. Therefore, in this study, the influence of the FFF process on the mechanical performance of lattice structures is specifically linked to in-plane inter-bead damage.

**Supplementary Note 4. Differential Scanning Calorimetry (DSC)**

The glass transition temperatures of the beads and filament were determined by DSC Seiko Exstar 6000 instrument. Samples were heated at a rate of 10 ºC/min from room temperature (25 ºC) to 250 ºC with a constant nitrogen flow of 50 mL/min. The $T_g$ values were calculated based on the midpoint between extrapolated heat flow as is shown in Supplementary Fig. 8. The amorphous nature of PEI was confirmed both before and after the additive manufacturing process because no exothermic transition due to crystallization was observed for all samples. Just a slight change of $T_g$, from 182°C to 177°C, was displayed after the FFF process suggesting that the polymer microstructures was slightly affected by the manufacturing process. The $T_g$ value of 182°C for the PEI filament tested confirms that it is Ultem$^{TM}$ 9085.

**Supplementary Note 5. FE validation of bulk material properties**

Supplementary Fig. 6 illustrates the FE validation of the compressive characteristics of FFF-printed bulk material fabricated. The numerical model consistently captures both elastic and plastic behaviour, although a slight disparity is noted in the yielding point and peak stress when compared to experimental data. Notably, the experimental results display a smoother response than the FE model, attributed to the inhomogeneous deformation of the FFF-printed cylindrical sample during significant deformation. This is a consequence of porosities between beads and inter-bead damage. In contrast, the numerical results are associated with an idealized cylinder.

Additionally, Supplementary Fig. 7 depicts the FE validation of the triaxial behaviour of FFF-printed bulk material. The numerical model demonstrates consistent results in terms of stress-strain curves and deformation. A comparison between the DIC strain map and the FE counterpart in Supplementary Fig. 7 reveals a favourable agreement, emphasizing the validation's robustness.

**Supplementary Note 6. FE validation of hexagonal lattice structure**

The numerical predictions of in-plane compression behaviour of hexagonal lattice structure with $\bar{\rho}$ = 30% are compared with the experimental results in Supplementary Fig. 13, showing the characteristic engineering stress-strain curves, deformation maps and failures at different stages. In this case, both the FE and the experimental results share the same collapse mechanism in the plateau regime: plastic yielding with subsequent fractures induced by arches of the cells[3]. The main difference between numerical and experimental results is the shape of the collapse of bending cells as highlighted by the ellipses in Supplementary Fig. 13 at stage B. The *D-P model with BC* predicts brittle fractures of the ligaments once the hexagonal cell is no longer able to withstand compression loading. On the contrary, experimental results and *D-P model with BC & Debonding* show collapsed hexagonal cells in which the ligaments remain connected. These ligaments have extended interlayer delamination that enables beads to slide instead of breaking. The experimental stress-strain curve exhibits a smooth transition from the elastic to plateau regime coupled with a slight hardening up to the first fracture, showing initial collapse stress 3 times lower than that predicted by the *D-P model with BC*. Although the collapse mechanism is similar, the inhomogeneous deformations are again limited to the bead-bead interfaces. The progressive failure of the cohesive zone involves hexagonal cell crushing at quite constant stress. The cell walls become flexible enough to cause extended collapse bands to occur at the cell arches as highlighted by the ellipses in Supplementary Fig. 13 at stage A.

**Tables of Content**

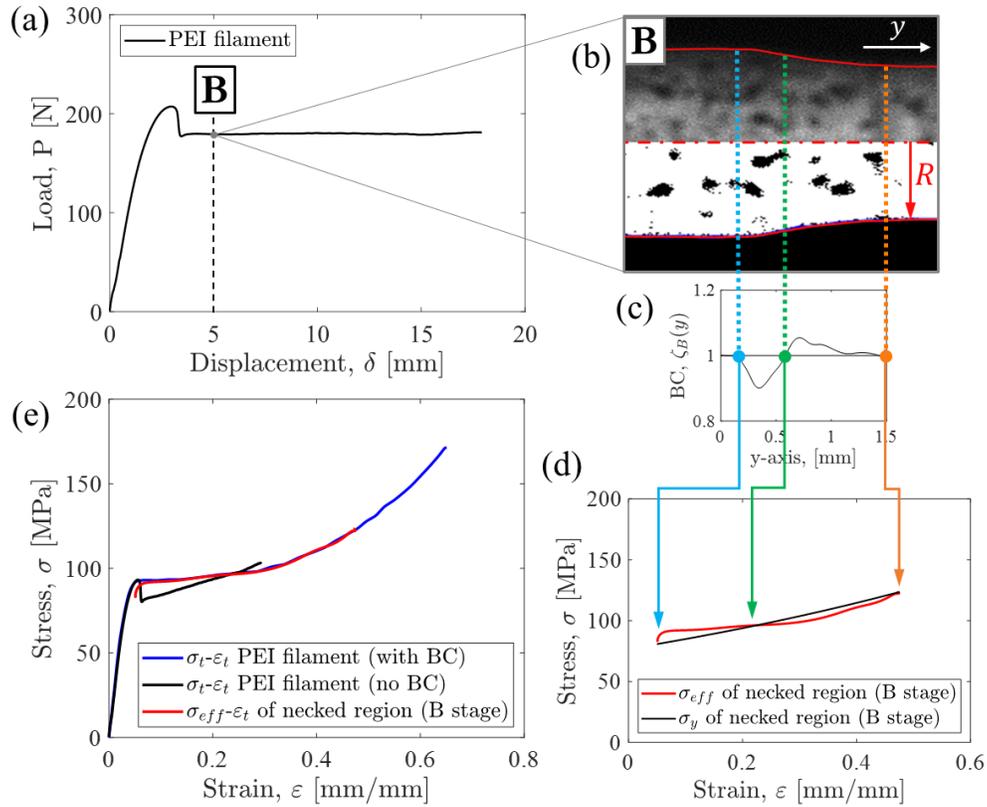

Supplementary Fig. 1: Identification of true stress-strain curve of PEI filament though G'Sell procedure and Bridgman correction. (a) Tensile load-displacement response of PEI filament. (b) Necked region exhibited by filament at stage B with subsequent image post-processing for filment profile identification (c) Bridgman's correction parameter computed along the filament axis. (d) Comparison between $\sigma_{eff}$ and $\sigma_t$ in the necked region. (e) Effect of Bridgman's correction on characteristic stress-strain curves of PEI filament.

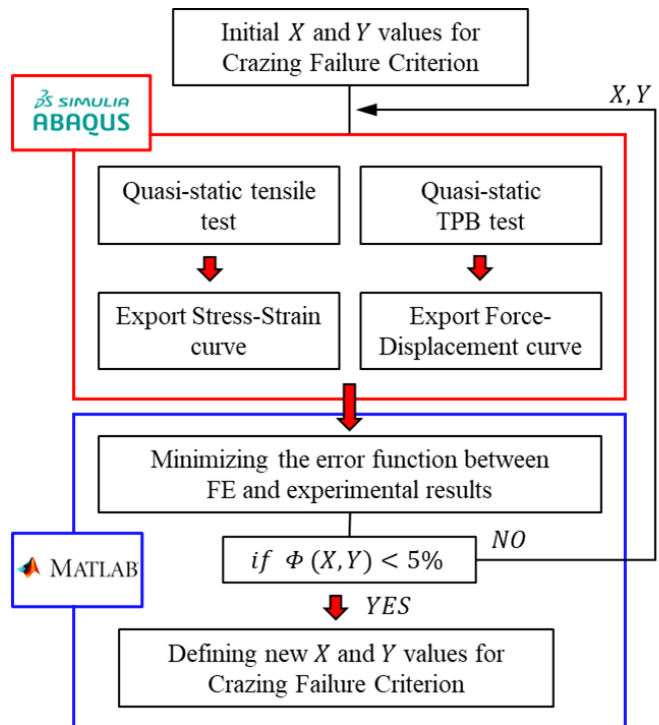

Supplementary Fig. 2: Flowchart for inverse identification procedure by FEMU.

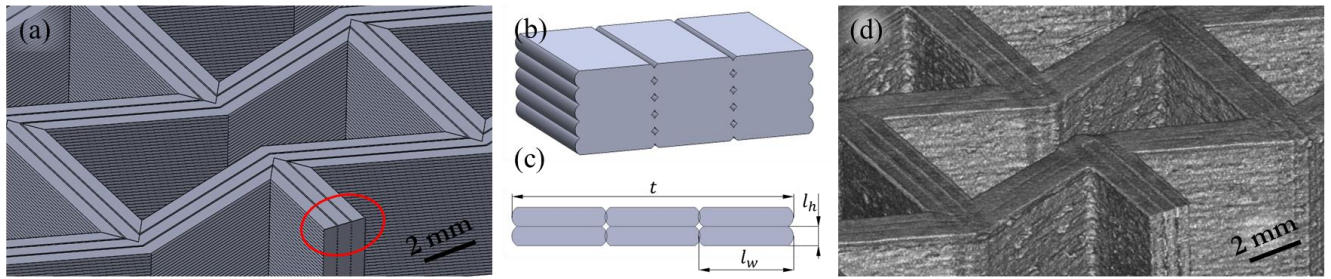

Supplementary Fig. 3: Bead-bead interfaces in FFF-printed lattice structures: (a) CAD model of re-entrant lattice structures with $\bar{\rho}$ =40%, (b) Portion of re-entrant lattice structure cell wall pointed out by red ellipse. (c) Geometric details of the beads within the cell walls of lattice structures ($t$ is the wall thickness, $l_w$ is the bead width and $l_h$ is the bead\layer hight) (d) Volume reconstruction of re-entrant lattice structures with $\bar{\rho}$=40% though micro-CT analysis.

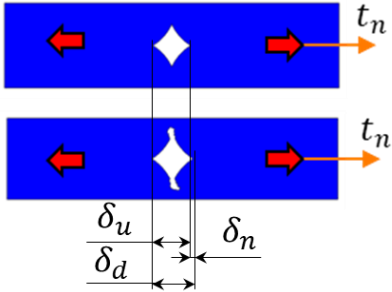

Supplementary Fig. 4: Determining separation values during numerical testing of RVE-B.

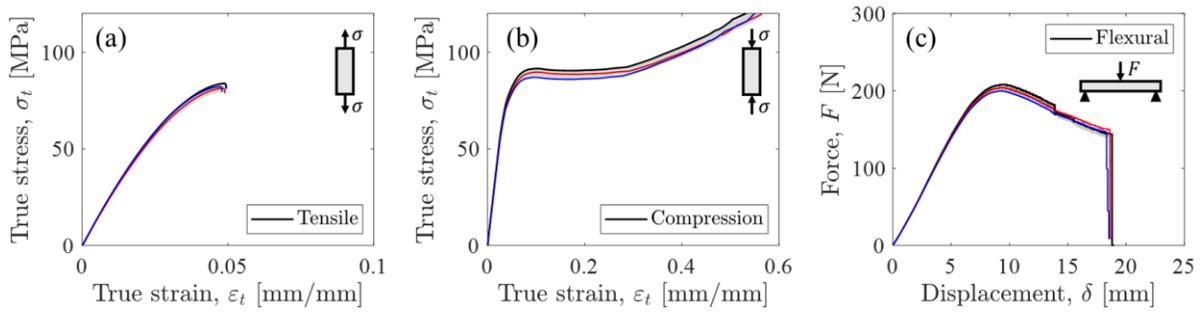

Supplementary Fig. 5: Quasi-static responses of FFF-printed PEI bulk samples: (a) under tension, (b) under compression and (c) under bending.

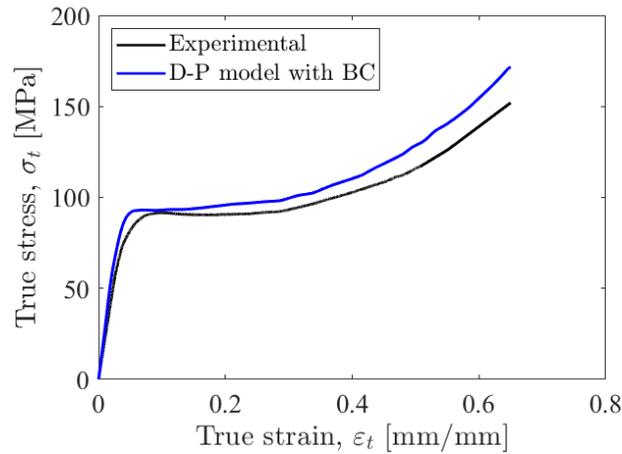

Supplementary Fig. 6: FE validation of compression test.

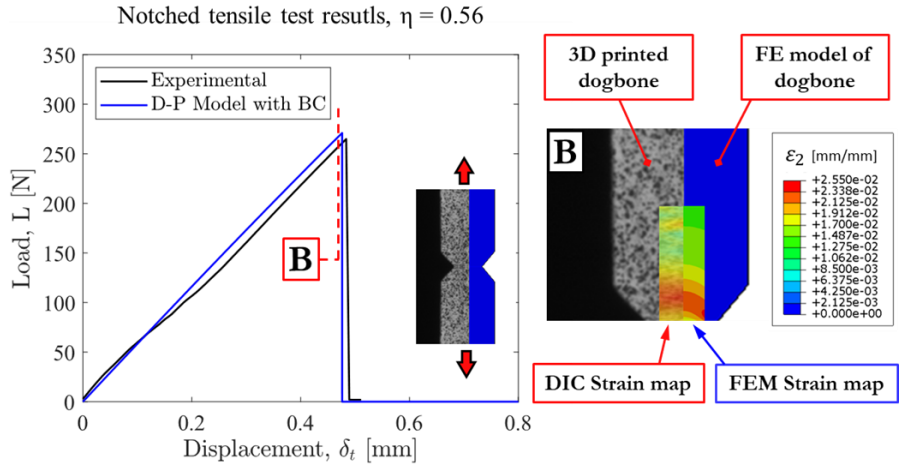

Supplementary Fig. 7: FE validation of notched test

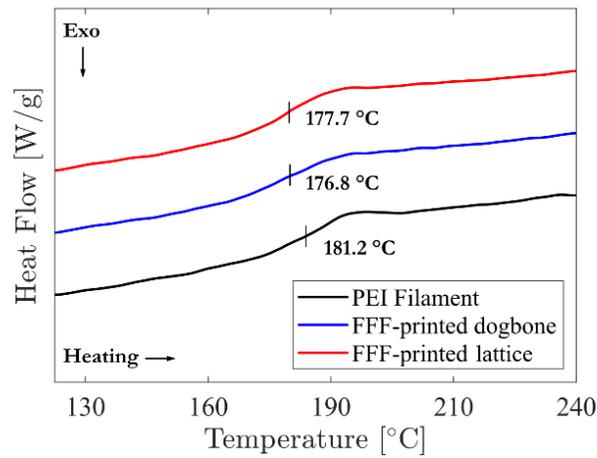

Supplementary Fig. 8: Differential Scanning Calorimetry (DSC) thermograms for PEI filament, FFF-printed dogbone sample and re-entrant lattice structure.

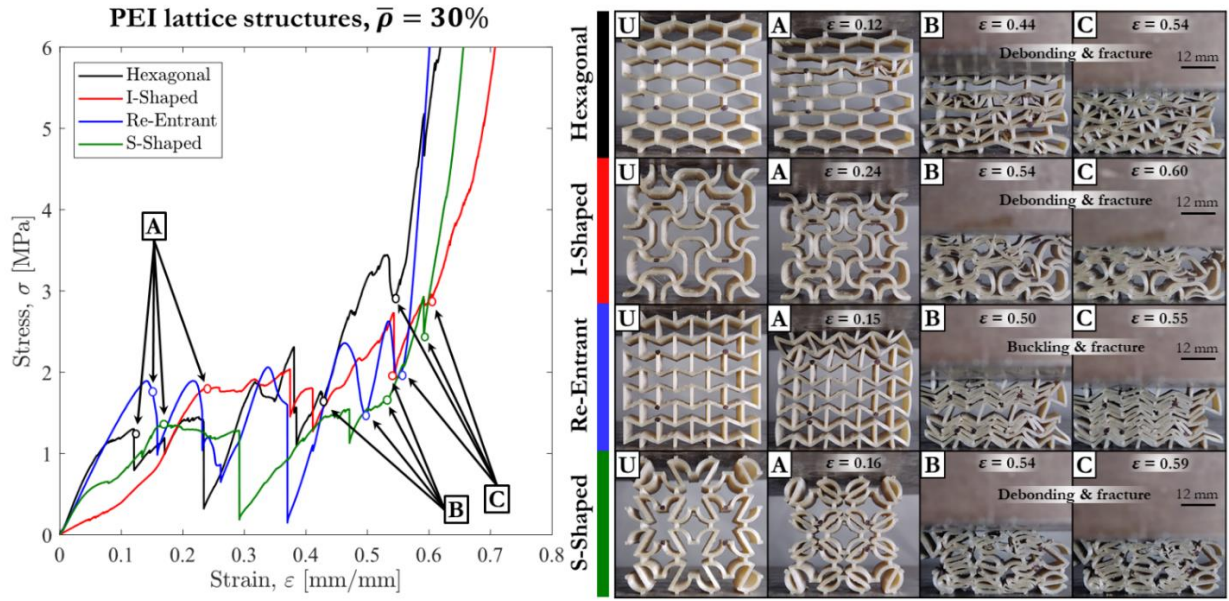

Supplementary Fig. 9: In-plane quasi-static compression behaviour of PEI honeycombs with $\bar{\rho} = 30\%$. Characteristic engineering stress-strain response, deformation maps and failures at various stages: U (Undeformed), A (Initial collapse stress), and B (Onset of densification), and C (Densification).

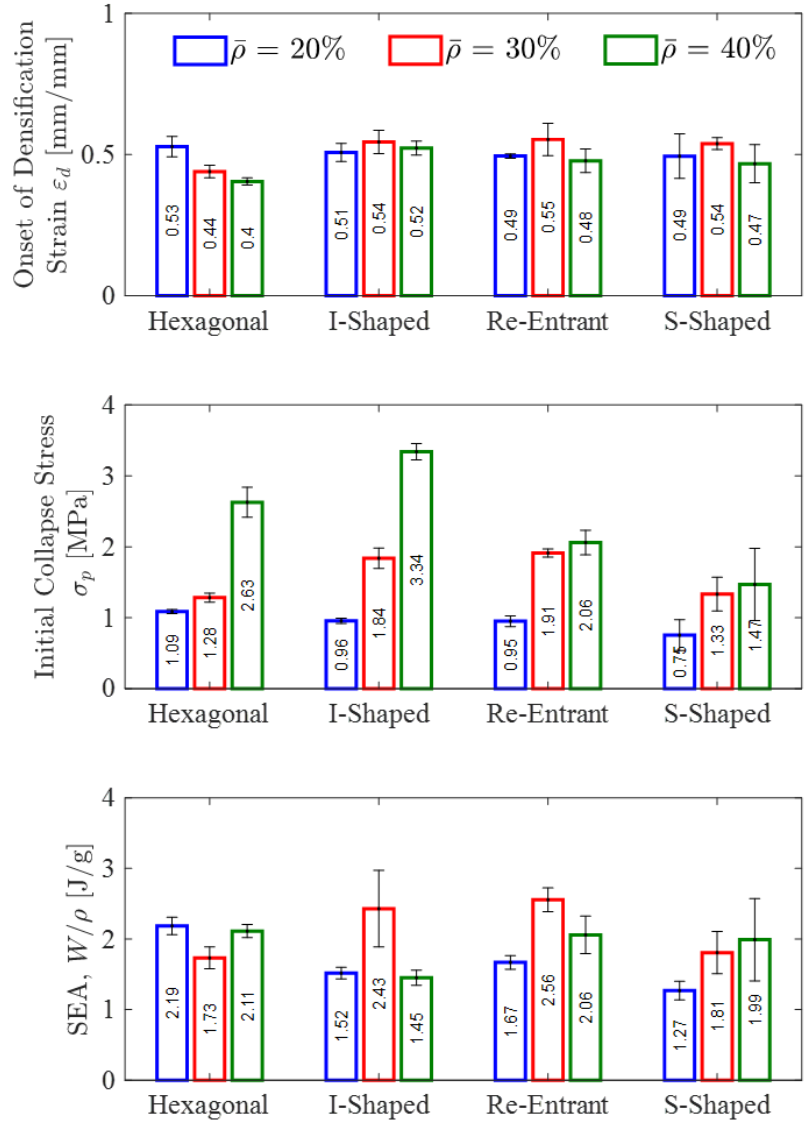

Supplementary Fig. 10: In-plane mechanical characteristics of FFF-printed PEI lattices under compression: Onset of densification strain, initial collapse stress and Specific Energy Absorption (SEA). The numerical values are summarised in Table 3 including Young's modulus, and Poisson's ratio.

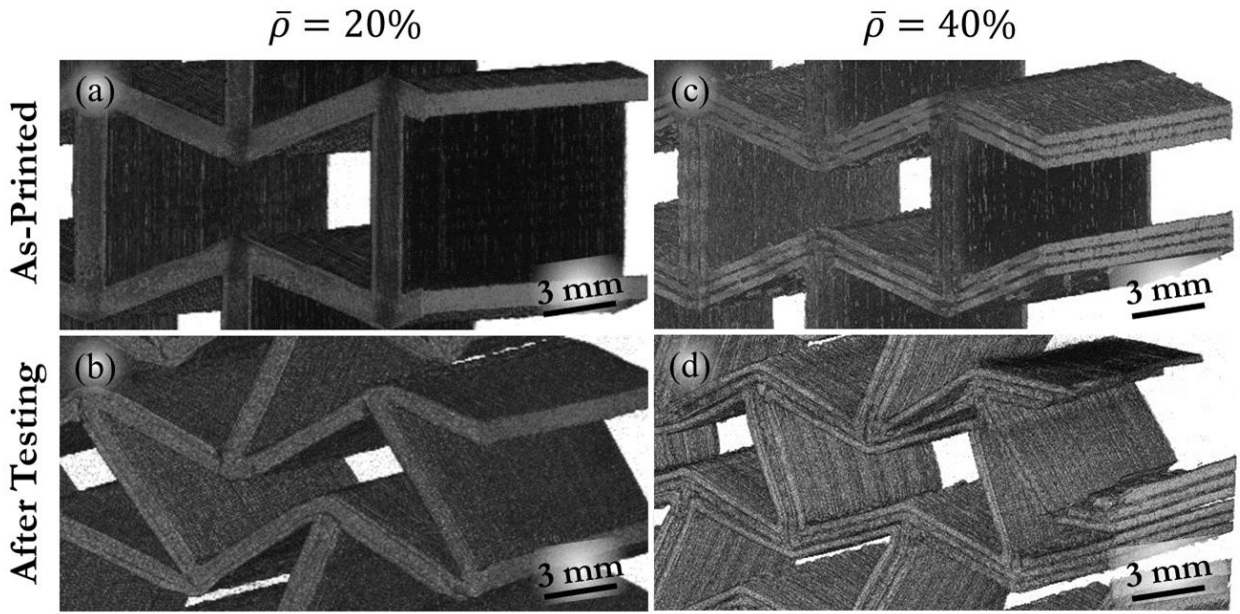

Supplementary Fig. 11: Volume reconstruction from micro-CT Analysis of Re-Entrant structures: (a) $\bar{\rho}$ = 20% - before compression tests. (b) $\bar{\rho}$ = 20% - after compression tests. (c) $\bar{\rho}$ = 40% - before compression tests. (d) $\bar{\rho}$ = 40% - after compression tests

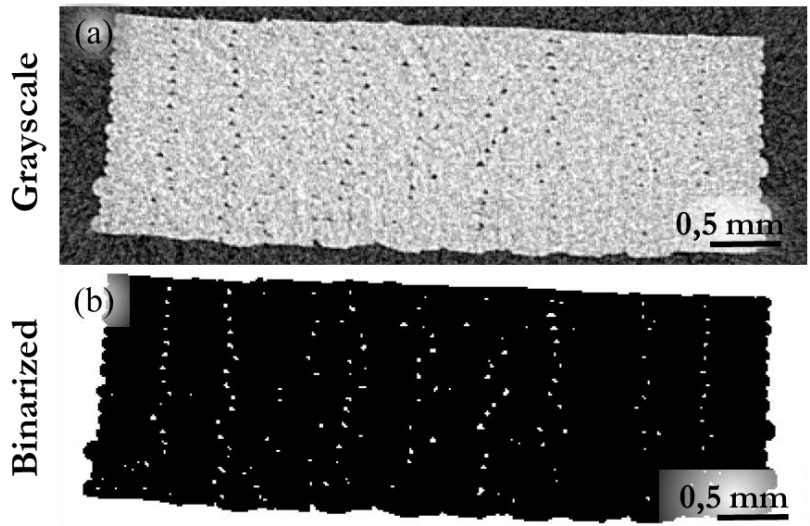

Supplementary Fig. 12: Porosity inside the FFF-printed PEI printed dogbone sample. (a) Grayscale image of dogbone sample cross-section. (b) Binarized image for porosity identification.

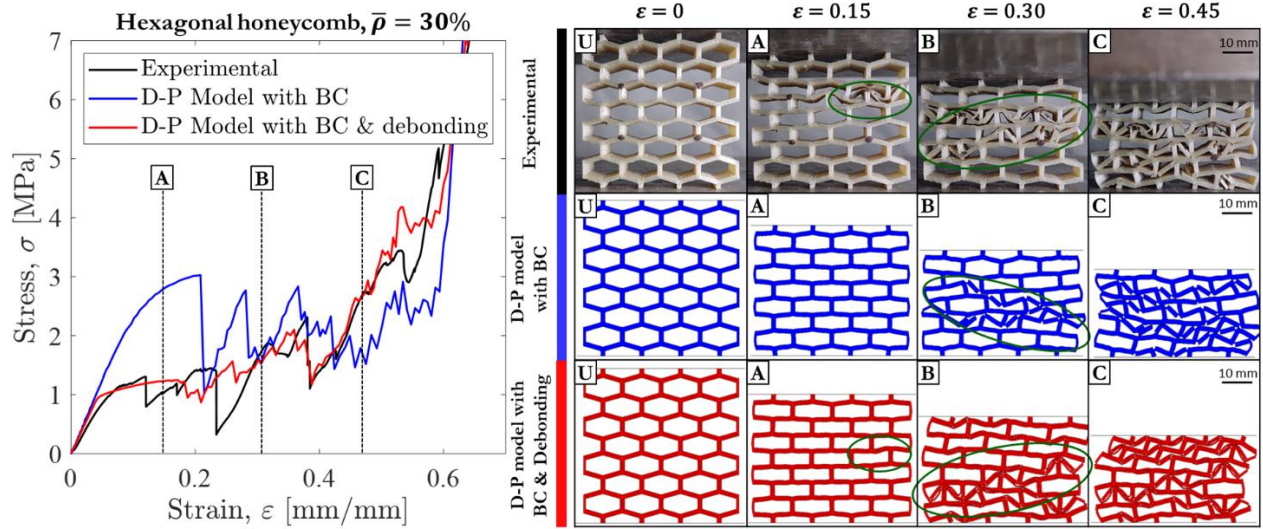

Supplementary Fig. 13: Experimental vs FE predictions of in-plane compression behaviour of hexagonal lattices with $\bar{\rho}$ = 30%. Characteristic engineering stress-strain response and deformation maps at various stages including failure modes are shown: U (Undeformed), A ($\varepsilon$ = 15%), B ($\varepsilon$ = 30%), and C ($\varepsilon$ = 45%).

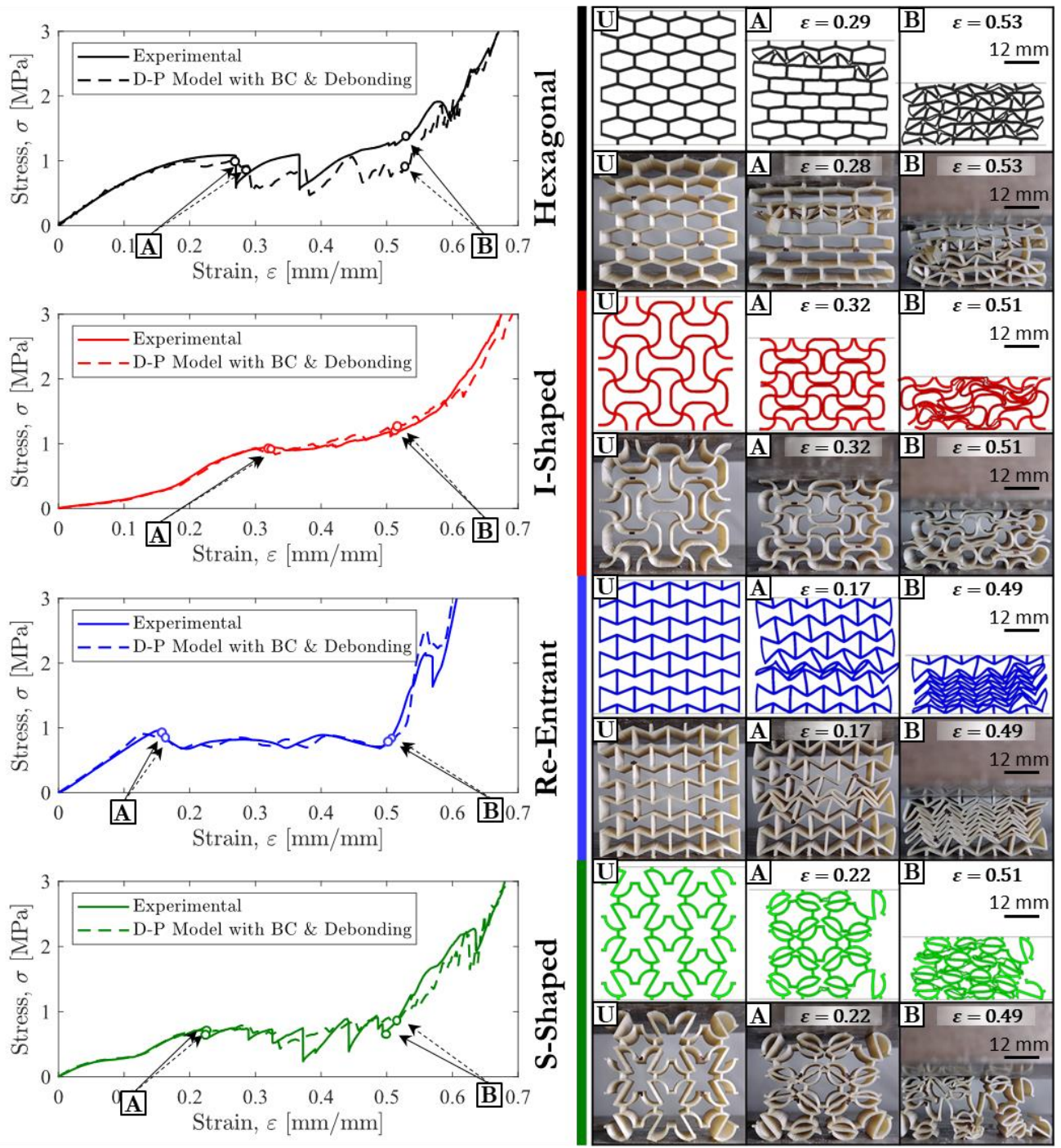

Supplementary Fig. 14: Numerical validation of in-plane compression behaviour of PEI honeycombs with $\bar{\rho} = 20\%$. Characteristic engineering stress-strain response, deformation maps and failures at various stages: U (Undeformed), A (Initial collapse stress), and B (Onset of densification).

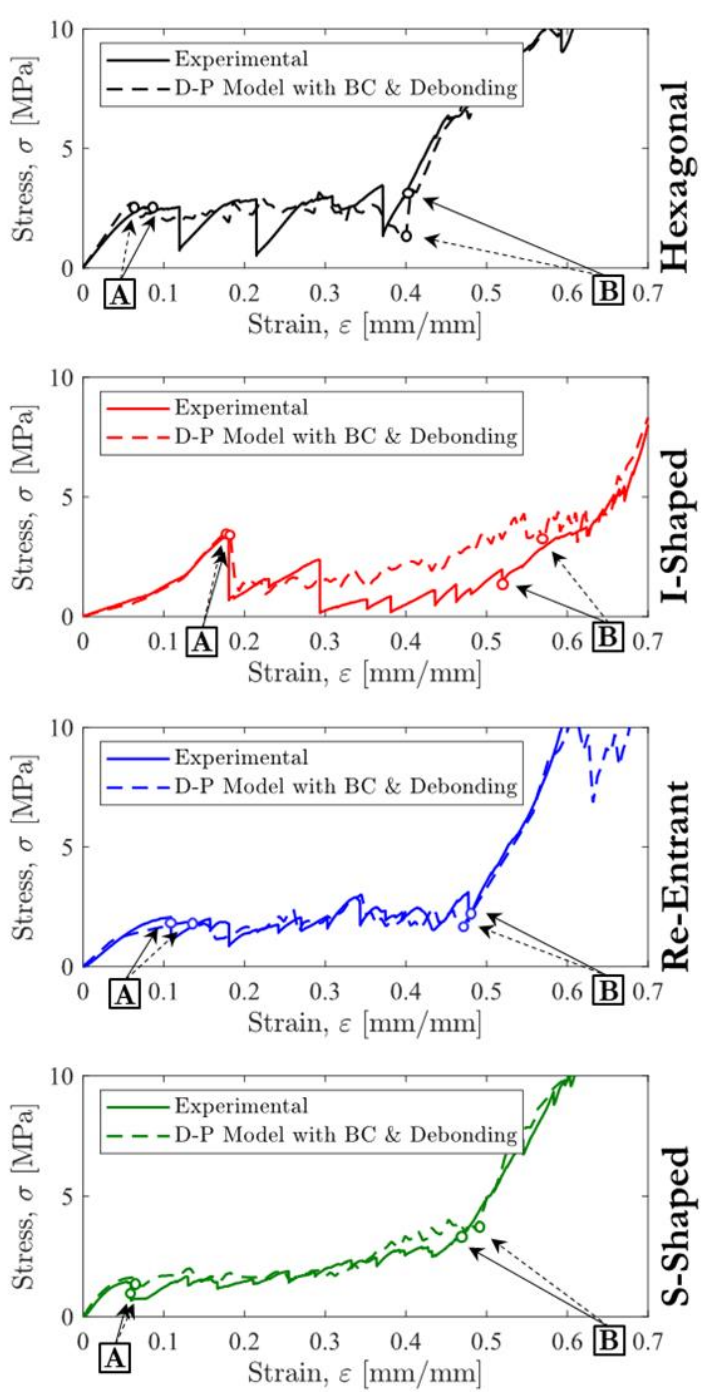
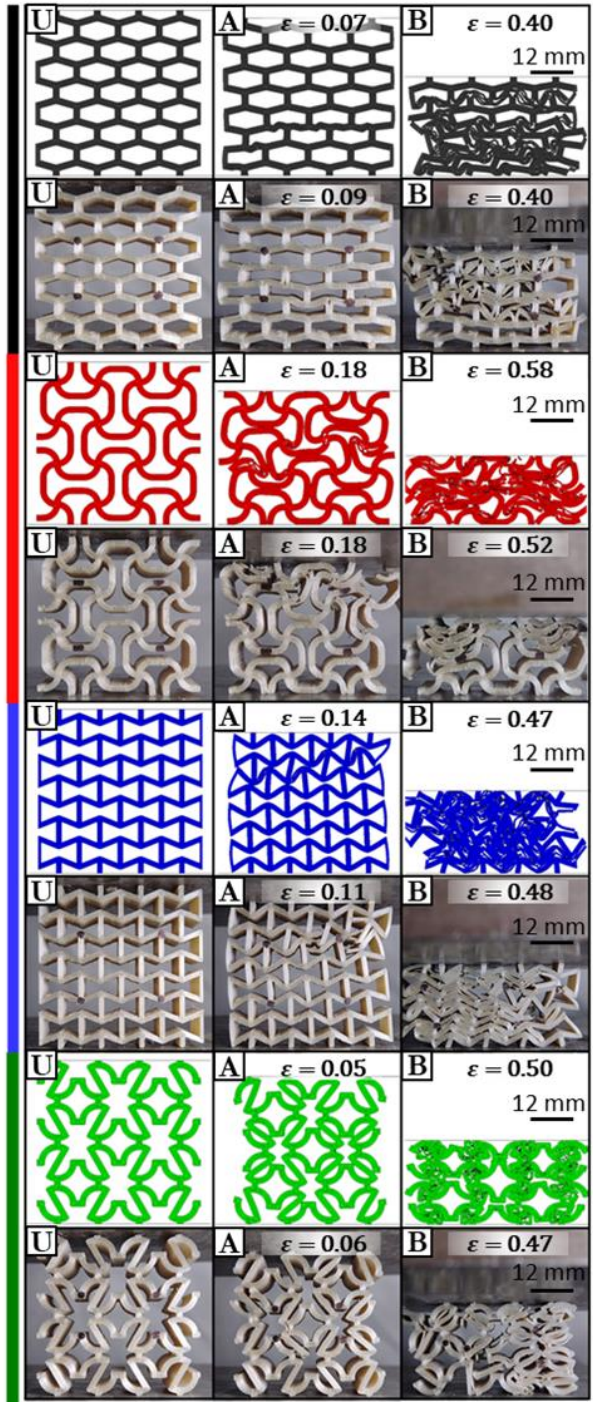

Supplementary Fig. 15: Numerical validation of in-plane compression behaviour of PEI honeycombs with $\bar{\rho}$ = 40%. Characteristic engineering stress-strain response, deformation maps and failures at various stages: U (Undeformed), A (Initial collapse stress), and B (Onset of densification).

Supplementary Table 1: FFF printing parameters for Apium PEI (Ultem$^{TM}$ 9085).

| Layer height [mm] | Nozzle diameter [mm] | Printing speed [mm/min] | Extruder temperature [°C] | Bed temperature [°C] | Infill percentage [%] | G-code software |
|---|---|---|---|---|---|---|
| 0.1 | 0.4 | 1000 | 370 | 120 | 100 | Simplify3D |

Supplementary Table 2: Architectural parameters of FFF-printed lattices.

| Unit cell topology | $\bar{\rho}$ [%] | $t$ [mm] | $a$ [mm] | $\theta$ [°] |
|---|---|---|---|---|
| Hexagonal | 20 | 0.90 | 4.5 | 60 |
| Hexagonal | 30 | 1.40 | 4.5 | 60 |
| Hexagonal | 40 | 1.90 | 4.5 | 60 |
| I-Shaped | 20 | 0.90 | 2 | 90 |
| I-Shaped | 30 | 1.37 | 2 | 90 |
| I-Shaped | 40 | 1.81 | 2 | 90 |
| Re-Entrant | 20 | 0.80 | 8.5 | 110 |
| Re-Entrant | 30 | 1.10 | 8.5 | 110 |
| Re-Entrant | 40 | 1.50 | 8.5 | 110 |
| S-Shaped | 20 | 0.90 | 4.5 | 120 |
| S-Shaped | 30 | 1.33 | 4.5 | 120 |
| S-Shaped | 40 | 1.79 | 4.5 | 120 |

**Supplementary References**